\documentclass[12pt]{article}

\usepackage{amsfonts,amssymb,amsmath}
\usepackage{latexsym}
\usepackage{graphicx}
\usepackage[dvips]{epsfig}
\usepackage{natbib}
\usepackage{epic,eepic}
\usepackage{appendix}
\usepackage{subfigure}

\usepackage{lineno}
\linenumbers

\def\RR{{\rm I\kern-.15em R}}

\newcommand{\xb}{\mbox{\boldmath$x$}}
\newcommand{\yb}{\mbox{\boldmath$y$}}

\newcommand{\Hb}{\mbox{\boldmath$H$}}
\newcommand{\Mb}{\mbox{\boldmath$M$}}
\newcommand{\Pb}{\mbox{\boldmath$P$}}
\newcommand{\Qb}{\mbox{\boldmath$Q$}}
\newcommand{\Rb}{\mbox{\boldmath$R$}}

\newcommand{\Sigmab}{\mbox{\boldmath$\Sigma$}}
\newcommand{\epsilonb}{\mbox{\boldmath$\epsilon$}}
\newcommand{\etab}{\mbox{\boldmath$\eta$}}

\begin{document}

%##############################
%  TITLE & AUTHORS
%##############################

\title{Particle Kalman Filtering: A Nonlinear Bayesian Framework for Ensemble
  Kalman Filters}

\author{{\Large Ibrahim Hoteit$^{1,}$\footnote{{\it Email}:
      ibrahim.hoteit@kasut.edu.sa;
      {\it Tel}: +966-544700033}, Xiaodong Luo$^1$, and Dinh-Tuan Pham$^2$}\vspace{2cm}\\
  $^1$King Abdullah University of Sciences and Technology, Thuwal, KSA \\
  $^2$Centre National de la Recherche Scientifique, Grenoble, France\vspace{2cm}\\
  Submitted to {\it Monthly Weather Review}\vspace{0.7cm}}

\date{\today}

\maketitle

%##############################
%  ABSTRACT
%##############################

\thispagestyle{empty}
\newpage
\setcounter{page}{0}

\begin{abstract}
  This paper investigates an approximation scheme of the optimal
  nonlinear Bayesian filter based on the Gaussian mixture
  representation of the state probability distribution function. The
  resulting filter is similar to the particle filter, but is different
  from it in that, the standard weight-type correction in the particle
  filter is complemented by the Kalman-type correction with the
  associated covariance matrices in the Gaussian mixture. We show that
  this filter is an algorithm in between the Kalman filter and the
  particle filter, and therefore is referred to as the particle Kalman
  filter (PKF).

  In the PKF, the solution of a nonlinear filtering problem is
  expressed as the weighted average of an ``ensemble of Kalman
  filters'' operating in parallel. Running an ensemble of Kalman
  filters is, however, computationally prohibitive for realistic
  atmospheric and oceanic data assimilation problems. For this reason,
  we consider the construction of the PKF through an ``ensemble'' of
  ensemble Kalman filters (EnKFs) instead, and call the implementation
  the particle EnKF (PEnKF).  We show that different types of the
  EnKFs can be considered as special cases of the PEnKF. Similar to
  the situation in the particle filter, we also introduce a
  re-sampling step to the PEnKF in order to reduce the risk of weights
  collapse and improve the performance of the filter. Numerical
  experiments with the strongly nonlinear Lorenz-$96$ model are
  presented and discussed.
\end{abstract}

\newpage

%##############################
%  INTRODUCTION
%##############################

\section{Introduction}

Estimating the state of the atmosphere and the ocean has long been one
of the main goals of modern science. Data assimilation, which consists
of combining data and dynamical models to determine the best possible
estimate of the state of a system, is now recognized as the best
approach to tackle this problem \citep{Ghil1991-data}.  The strongly
nonlinear character of the atmospheric and oceanic models, combined
with their important computational burden, makes data assimilation in
these systems quite challenging.

Based on the Bayesian estimation theory, the optimal solution of the
nonlinear data assimilation problem can be obtained from the optimal
nonlinear filter (ONF) \citep{Doucet2001-sequential}. This involves the
estimation of the conditional probability distribution function ({\it
  pdf}) (not necessarily Gaussian) of the system state given all
available measurements up to the estimation time. Knowledge of the
state {\it pdf} allows determining different estimates of the state,
such as the minimum variance estimate or the maximum a posteriori
estimate \citep{Todling1999-estimation}. The ONF recursively operates
as a succession of a correction (or analysis) step at measurement
times to correct the state (predictive) {\it pdf} using the Bayes'
rule, and a prediction step to propagate the state (analysis) {\it
  pdf} to the time of the next available observation. Although
conceptually simple, the numerical implementation of the optimal
nonlinear filter can be computationally prohibitive, even for systems
with few dimensions \citep{Doucet2001-sequential}. Its use with atmospheric and
oceanic data assimilation problems is therefore not possible because
of the huge dimension of these systems.

In recent years, two approximation schemes of the ONF have attracted
the attention of researchers for their potentials to tackle nonlinear
and non-Gaussian data assimilation problems. One is based on the
point-mass representation (mixture of Dirac functions) of the state
{\it pdf}, and leads to the celebrated particle filter (PF)
\citep{Doucet2001-sequential,Pham2001,Nakano2007-merging,vanLeeuwen-variance,VanLeeuwen2009}. The
other is based on the Gaussian mixture representation of the state
{\it pdf}, and results in a filter that is in between the Kalman
filter and the particle filter
\citep{Anderson-Monte,Bengtsson2003,Chen2000-mixture,Hoteit2008,Luo2008-spgsf1,Sorenson-recursive},
as to be shown later. For this reason, we refer to this filter as the
particle Kalman filter (PKF).

In terms of computational efficiency, the particle filter needs to
generate large samples for a good approximation of the state {\it
  pdf}. In certain circumstances, in order to avoid weights collapse,
the number of samples needs to scale exponentially with the dimension
of the system in assimilation \citep{Bengtsson2008}, which may be
infeasible for high-dimensional systems \citep{Snyder2008}. On the
other hand, in some comparison studies
\citep{Han-evaluation,Nakano2007-merging}, it has been reported that
the ensemble Kalman filter (EnKF) and its variants
\citep{Anderson-ensemble,Bishop-adaptive,Burgers-analysis,Evensen-sequential,Evensen-assimilation,Houtekamer1998,Whitaker-ensemble}
can achieve lower estimation errors than the particle filter given a
small ensemble size. To save space, in this paper we confine ourselves
to the PKF, and make performance comparison only between the PKF and
the EnKF.

Using a Gaussian mixture representation of the state {\it pdf}, the
resulting PKF consists of an ensemble of parallel nonlinear Kalman
filters \citep{Hoteit2008,Luo2008-spgsf1}. Different variants of the
Kalman filter (KF), including the extended Kalman filter
\citep{Chen2000-mixture,Sorenson-recursive}, the reduced-rank Kalman
filter \citep{Hoteit2008,Luo2008-spgsf1}, the EnKF
\citep{Anderson-Monte,Bengtsson2003}, can be used to construct the
PKF. The focus of this paper is to investigate the PKF that is
constructed by an ensemble of parallel EnKFs. Common to all the
implementations of the PKF, the mixture of normal distributions (MON)
-- a more general {\it pdf} representation than the single Gaussian
{\it pdf} approximation in the EnKF -- can be used to tackle
nonlinearity and non-Gaussianity in data assimilation. On the other
hand, choosing the EnKF to construct the PKF is based on the
consideration of computational efficiency, since the EnKF itself is a
very efficient algorithm for data assimilation in high dimensional
systems. In this regard, this work is very similar to the earlier
works of \citet{Anderson-Monte} and \citet{Bengtsson2003}, but is
different from them mainly in the following aspect.

In \citet{Anderson-Monte} and \citet{Bengtsson2003}, the PKF was
constructed without a re-sampling step. As a result, the PKF may
suffer from weights collapse as in the particle filter. To overcome
this problem, \cite{Bengtsson2003} considered a hybrid of the EnKF and
the PKF, which, however, involves the computation of the inverses of
sample covariance matrices in the ``global-to-local'' adjustments. In
doing so, it is not only computationally intensive, but also
encounters singularities in computing the inverses when the ensemble
size is smaller than the system dimension, such that the sample
covariances themselves are rank deficient. Therefore, it is not clear
how the hybrid scheme in \cite{Bengtsson2003} can be applied to the
scenario with the ensemble size smaller than the system dimension. For
the implementation of the PKF scheme in this work, we introduce a
re-sampling step similar to those in \cite{Musso-improving-2001} and
\citet{Stavropoulos-improved-2001} to tackle weights collapse. Our
experience shows that, with this re-sampling step, the PKF becomes
much more stable and can conduct data assimilation in the small
ensemble scenario, as to be demonstrated through the numerical
experiments presented in this work.

As may be of particular interest for the ensemble filtering community,
we will show that different EnKFs can be considered as special cases
of the PEnKF following our implementation. This point of view allows
for a better understanding of the EnKFs' behaviors and/or their
differences.

The paper is organized as follows. The optimal nonlinear filter is
first described in section \ref{optfilter}. The PKF and its ensemble
implementation are discussed in section \ref{pkf}. Results of
numerical experiments with the Lorenz-$96$ model are presented in
section \ref{experiments}. A summary of the main results and a general
discussion on the potential of the PEnKF for tackling realistic
atmospheric and oceanic data assimilation problems concludes the paper
in section \ref{discussion}.

%##############################
% THE OPTIMAL NONLINEAR FILTER
%##############################

\section{The Optimal Nonlinear Filter}
\label{optfilter}

Starting from a random initial condition with a known probability
density function, the optimal nonlinear filter provides the
conditional density function of the system state given all available
measurements up to the estimation time. To describe the algorithm of
the optimal nonlinear filter, consider the nonlinear stochastic
discrete-time dynamical system
\begin{eqnarray}
\label{model}
\xb_k &=& \Mb_k(\xb_{k-1}) + \etab_k,\\
\label{obs}
\yb_k &=& \Hb_k(\xb_k) + \epsilonb_k,
\end{eqnarray}
where $\xb_k$ is the state vector (to be estimated), of dimension $n$,
$\yb_k$ is the observation vector, of dimension $p$, $\Mb_k$ and
$\Hb_k$ are two continuously differentiable maps from $\RR^n$ to
$\RR^n$ and from $\RR^n$ to $\RR^p$ respectively representing the
transition and the observational operators, and $\etab_k$ and
$\epsilonb_k$ denote the dynamical and the observational noise,
respectively. We assume that $\etab_k$ and $\epsilonb_k$ are Gaussian
with zero mean and non-singular covariance matrices $\Qb_k$ and
$\Rb_k$, respectively, and are independent of the system state at any
time instant. Under this setting, the dynamical system
Eq.~(\ref{model}) is Markovian.

The optimal nonlinear filter recursively operates with a succession of
prediction and correction steps as summarized below. The reader is
referred to \cite{Doucet2001-sequential} for an extensive description of the
filter. To simplify the notation, $\yb_{1:k}$ is defined as a
shorthand for the set of all observations $\yb_1, \dots, \yb_k$ up to
and including time $t_k$. Let $p^f_k(~\cdot \mid\yb_{1:k-1})$ be the
conditional (predictive) {\it pdf} of $\xb_k$ given $\yb_{1:k-1}$ and
$p^a_k(~\cdot \mid \yb_{1:k})$ be the conditional (analysis) {\it pdf}
of $\xb_k$ given $\yb_{1:k}$, both determined at time $t_k$. The
filter steps are described as follows.

\begin{enumerate}

\item[$\bullet$] {\underline {\it Prediction step}:} Given the
analysis {\it pdf} $p^a_{k-1}(~\cdot \mid \yb_{1:k-1})$ at time
$t_{k-1}$, the predictive {\it pdf} $p^f_k(~\cdot \mid \yb_{1:k-1})$ is
obtained by integrating $p^a_{k-1}(~\cdot \mid \yb_{1:k-1})$ with the
model (\ref{model}) to the time of the next available observation
$t_k$. Under the assumptions made on the model noise $\etab_k$, the
likelihood function for the state vector $\xb_{k-1}$ to transit to $\xb_{k}$ at the next time instant is described by the Gaussian {\it pdf} $N \left(\xb_{k}: \Mb_k(\xb_{k-1}), \Qb_k\right)$, where $N \left(\xb: \mathbf{\mu}, \Sigmab\right)$ denotes the Gaussian {\it pdf} with mean $\mathbf{\mu}$ and covariance $\Sigmab$. Thus,
\begin{eqnarray}
\label{e:predicden}
p^f_k(\xb_{k} \mid \yb_{1:k-1}) &=& \int_{\RR^n} N \left(\xb_{k}:
\Mb_k(\xb_{k-1}) , \Qb_k\right) p^a_{k-1}(\xb_{k-1} \mid \yb_{1:k-1}) d\xb_{k-1}.
\end{eqnarray}

\item[$\bullet$] {\underline {\it Correction step}:} After a new
observation $\yb_k$ has been made, the analysis {\it pdf} $p_k^a (~\cdot
\mid \yb_{1:k})$ at time $t_k$ is updated from $p_k^f (~\cdot
\mid \yb_{1:k-1})$ using Bayes' rule, i.e.,
\begin{eqnarray}
\label{e:conden}
p^a_k(\xb_{k} \mid \yb_{1:k}) &=& \frac{1}{b_k}p^f_k(\xb_{k} \mid
\yb_{1:k-1}) N \left(\yb_k: \Hb_k(\xb_{k}), \Rb_k\right).
\end{eqnarray}
The analysis {\it pdf} is therefore obtained by multiplying the
predictive {\it pdf} by the observation likelihood function $N \left(\yb_k: \Hb_k(\xb_{k}), \Rb_k\right)$, and then being normalized by $b_k
= \int_{\RR^n} p^f_k(\xb_{k} \mid \yb_{1:k-1}) N\left(\yb_k:
H_k(\xb_{k}) , R_k\right) d\xb_{k}$.

\end{enumerate}

While the expressions of the state {\it pdf}s can be obtained
conceptually, determining the exact values of them at each point of
the state space is practically infeasible in high dimensional systems
\citep{Doucet2001-sequential}. For instance, the determination of the predictive
{\it pdf} requires the evaluation of the model $\Mb_k(\xb)$ for a
prohibitively large number of $\xb$, given that one single evaluation
might already be computationally very expensive in realistic
atmospheric and oceanic applications. %\citep{Snyder2008}

%##############################
%  The Particle Kalman Filter
%##############################

\section{The Particle Ensemble Kalman Filter}
\label{pkf}

\subsection{Particle Kalman Filtering and Its Ensemble Implementation}

Given $N$ independent samples $\xb^1, \ldots, \xb^N$ from a
(multivariate) density $p$, an estimator $\hat{p}$ of $p$ can be
obtained by the kernel density estimation method
\citep{Silverman1986}, in the form of a mixture of $N$ Gaussian {\it
  pdf}s:
\begin{eqnarray}
\hat{p}(\xb) &=& \frac{1}{N}\sum_{i=1}^N N(\xb: \xb^i , \Pb),
\end{eqnarray}
where $\Pb$ is a positive definite matrix. Inspired from this
estimator, the particle Kalman filter (PKF) approximates the
conditional state {\it pdf}s in the optimal nonlinear filter by
mixtures of $N$ Gaussian densities of the form
\begin{eqnarray}
\label{approxmix}
p^s_k(\xb_k \mid \yb_{1:k}) &=& \sum_{i=1}^N w^i_k N(\xb_k: \xb^{s,i}_k, \Pb^{s,i}_k).
\end{eqnarray}
The subscript $s$ replaces $a$ at the analysis time and $f$ at the
prediction time. The parameters of the mixture are the weights
$w^i_k$, the centers of the distributions $\xb^{s,i}_k$, and the
covariance matrices $\Pb^{s,i}_k$. In particular, if $N=1$,
$p^s_k(\xb_k \mid \yb_{1:k})$ reduces to a single Gaussian {\it pdf},
so that the PKF reduces to the Kalman filter (KF) or its variants
trivially (a non-trivial simplification will also be discussed
below). Consequently, the KF and its variants can be considered
special cases of the PKF.

Two special cases of Eq.~(\ref{approxmix}) may be of particular
interest. In the first case, $\Pb^{s,i}_k \rightarrow \mathbf{0}$,
such that the Gaussian {\it pdf}s $N(\xb_k: \xb^{s,i}_k, \Pb^{s,i}_k)$
tend to a set of Dirac functions $\delta(\xb^{s,i}_k)$, with the mass
points at $\xb^{s,i}_k$. In this case, the Gaussian mixture
Eq.~(\ref{approxmix}) reduces to the Monte Carlo approximation used in
the particle filter \citep{Doucet2001-sequential}. In the second case, all
Gaussian {\it pdf}s $N(\xb_k: \xb^{s,i}_k, \Pb^{s,i}_k)$ have (almost)
identical centers and covariances, such that the Gaussian mixture
Eq.~(\ref{approxmix}) tends to a (single) Gaussian approximation, an
assumption often used in various nonlinear Kalman filters (including
the EnKF). In this sense, the PKF can be considered as a filter in
between the Kalman filter and the particle filter
\citep{Hoteit2008,Luo2008-spgsf1}.

The main procedures of the PKF are summarized as follows. Without loss
of generality, suppose that at time instant $k-1$, the analysis {\it
  pdf}, after a re-sampling step, is given by
$\tilde{p}_{k-1}(\xb_{k-1} \mid \yb_{1:k-1}) = \sum_{i=1}^N
\tilde{w}^i_{k-1} N(\xb_{k-1}: \mathbf{\theta}^{i}_{k-1},
\mathbf{\Phi}^{i}_{k-1})$. Then by applying Eq.~(\ref{e:predicden}) at
the prediction step, one obtains the background {\it pdf}, in terms of
a new MON
\begin{eqnarray}
\label{approxpred}
p^f_k(\xb_k \mid \yb_{1:k-1}) &\approx& \sum_{i=1}^N \tilde{w}^i_{k-1}
N \left(\xb_k: \hat{\xb}^{f,i}_{k} , \hat{\Pb}^{f,i}_{k}\right),
\end{eqnarray}
where $\hat{\xb}^{f,i}_{k}$ and $\hat{\Pb}^{f,i}_{k}$ are the
propagations of the mean $\mathbf{\theta}^{i}_{k-1}$ and the
covariance $\mathbf{\Phi}^{i}_{k-1}$ of the Gaussian component
$N(\xb_{k-1}: \mathbf{\theta}^{i}_{k-1}, \mathbf{\Phi}^{i}_{k-1})$
through the system model Eq.~(\ref{model}), respectively.

Given an incoming observation $\yb_k$, one applies
Eq.~(\ref{e:conden}) to update $p^f_k(\xb \mid \yb_{1:k-1})$ to the
analysis {\it pdf}, also in the form of an MON
\begin{eqnarray}
\label{densmix}
p^a_k(\xb_k \mid \yb_{1:k}) &=& \sum_{i=1}^N w^i_{k} N \big(\xb_k:
\hat{\xb}^{a,i}_{k}, \hat{\Pb}^{a,i}_{k}\big),
\end{eqnarray}
where $\hat{\xb}^{a,i}_{k}$ and $\hat{\Pb}^{a,i}_{k}$ are updated from
$\hat{\xb}^{f,i}_{k}$ and $\hat{\Pb}^{f,i}_{k}$ through the Kalman
filter or its variants, and the new weights
\begin{eqnarray}
\label{corrp}
w^i_k &=& \frac{\tilde{w}^i_{k-1} N \big(\yb_k: \Hb_k(\hat{\xb}^{f,i}_k) ,
\Sigmab^i_k \big)}{\sum_{j=1}^N \tilde{w}^i_{k-1} N \big(\yb_k: \Hb_k(\hat{\xb}^{f,i}_k), \Sigmab^i_k \big)} \, ,
\end{eqnarray}
where $\Sigmab^i_k$ is the innovation matrix. If evaluated through the
extended Kalman filter, $\Sigmab^i_k = {\bf H}^i_k \hat{\Pb}^{f,i}_k
({\bf H}_k^{i})^T + \Rb_k$, with ${\bf H}^i_k$ being the gradient of
$\Hb_k$ evaluated at $\hat{\xb}^{f,i}_k$. Alternatively, if evaluated
in the context of the EnKF, $\Sigmab^i_k$ can be expressed as the
covariance of the projected background ensemble onto the observation
space plus the observation covariance $\Rb_k$
\citep{Evensen-sequential,Whitaker-ensemble}. Finally, a re-sampling
step can be introduced to improve the performance of the PKF
\citep{Hoteit2008,Luo2008-spgsf1}, so that the analysis {\it pdf}
becomes $\tilde{p}_{k}(\xb_{k} \mid \yb_{1:k}) = \sum_{i=1}^N
\tilde{w}^i_k N(\xb_{k}: \mathbf{\theta}^{i}_{k},
\mathbf{\Phi}^{i}_{k})$. Such a re-sampling algorithm is presented in
the next section.

The PKF correction step can be interpreted as composed of two types of
corrections: a {\it Kalman-type correction} used to update
$\hat{\xb}^{f,i}_{k}$ and $\hat{\Pb}^{f,i}_{k}$ to
$\hat{\xb}^{a,i}_{k}$ and $\hat{\Pb}^{a,i}_{k}$, and a {\it
  particle-type correction} used to update the weights
$\tilde{w}^i_{k-1}$ to $w^i_{k}$. In the PKF, the Kalman correction
reduces the risk of weights collapse by allocating the estimates
$\hat{\xb}^{f,i}_k$ (whose projections onto the observation space) far
away from the observation $\yb_k$ relatively more weights than in the
particle filter \citep{Hoteit2008,VanLeeuwen2009}. Indeed,
Eq. (\ref{corrp}) has the same form as in the PF \citep{Doucet2001-sequential},
but uses the innovation matrices $\Sigmab^i_k$ to normalize the
model-data misfit, rather than $\Rb_k$. As $\Sigmab^i_k$ are always
greater than $\Rb_k$, the estimates that are close to the observation
will receive relatively less weights than in the PF, while those far
from the observation will receive relatively more weights. This means
that the support of the local predictive {\it pdf} and the observation
likelihood function will be more coherent than in the PF. Re-sampling
will therefore be needed less often, so that Monte Carlo fluctuations
are reduced.

The main issue with the PKF is the prohibitive computational burden
associated with running an ensemble of KFs, knowing that running a
Kalman filter (KF) or an extended KF in high dimensional systems is
already a challenge. To reduce computational cost, we use an ensemble
of EnKFs, rather than the KF or the extended KF, to construct the
PKF. We refer to this approach as the Particle Ensemble Kalman Filter
(PEnKF). In the PEnKF, the (analysis) ensembles representing the
Gaussian components are propagated forward in time to obtain a set of
background ensembles at the next assimilation cycle. Then for each
background ensemble, a stochastic or deterministic EnKF is used to
update the background ensemble to its analysis counterpart. This
amounts to simultaneously running a weighted ensemble of EnKFs, and
the final state estimate is the weighted average of all the EnKFs
solutions.

\subsection{A Re-sampling Algorithm}
\label{sec:re-sampling}

We adopt a re-sampling algorithm that combines those in
\citet{Hoteit2008,Luo2008-spgsf1,Pham2001}. The main idea is as
follows: Given a MON, we first employ an information-theoretic
criterion used in \citet{Hoteit2008} and \citet{Pham2001} to check if
it needs to conduct re-sampling. If there is such a need, we then
re-approximate the MON by a new MON, based on the criterion that the mean and covariance of the new MON match those of the original MON as far as possible \citet{Luo2008-spgsf1}.

More concretely, let $p \left( \mathbf{x} \right)$ be the {\it pdf} of
the $n$-dimensional random vector $\mathbf{x}$, expressed in terms of
an MON with $N$ Gaussian {\it pdf}s so that
\begin{equation} \label{eq:original_GMM}
p \left( \mathbf{x} \right) = \sum\limits_{i=1}^{N} w_i  N \left(\mathbf{x}: \mathbf{\mu}_i, \mathbf{\Sigma}_i\right) \, ,
\end{equation}
where $w_i$ are the set of normalized weights of the Gaussian {\it
  pdf}s $N \left(\mathbf{x}: \mathbf{\mu}_i, \mathbf{\Sigma}_i\right)$
with mean $\mathbf{\mu}_i$ and covariance $\mathbf{\Sigma}_i$,
satisfying $w_i \geq 0$ for $i=1,\dotsb,N$ and $\sum_{i=1}^{N}
w_i=1$. To decide whether to conduct re-sampling or not, the entropy
$E_w$ of the weights $w_i$ is computed, which reads
\citep{Hoteit2008,Pham2001}
\begin{equation}
E_w = - \sum\limits_{i=1}^{N} w_i \text{log} w_i \, .
\end{equation}
Ideally, when the distribution of the weights $w_i$ is uniform, which
yields the maximum weight entropy $E_w^{u}=\text{log} N$, there is no
need to conduct re-sampling. Thus, as a criterion, if $E_w$ is within
a certain distance $d$ to $E_w^u$, i.e.,
\begin{equation} \label{eq:resampling_criterion}
E_w^u - E_w = \text{log} N + \sum\limits_{i=1}^{N} w_i \text{log} w_i \leq d \, ,
\end{equation}
where $d$ is a user-defined threshold, then we choose not to conduct
re-sampling. In this work we set the threshold $d=0.25$ following
\citet{Hoteit2008}.

In case that there is a need to conduct re-sampling, we follow the
procedure similar to that in \citet{Luo2008-spgsf1}. Here the idea is
to treat re-sampling as a {\it pdf} approximation problem, in which we
seek a new MON
\begin{equation} \label{eq:new_GMM}
\tilde{p} \left( \mathbf{x} \right) = \dfrac{1}{q} \sum\limits_{i=1}^{q} N \left(\mathbf{x}: \mathbf{\theta}_i, \mathbf{\Phi}_i\right) \, ,
\end{equation}
with $q$ equally weighted Gaussian {\it pdf}s, to approximate the
original $p \left( \mathbf{x} \right)$ in
Eq.~(\ref{eq:original_GMM}). In approximation, we require that the
mean and covariance of $\tilde{p} \left( \mathbf{x} \right)$ be as
close as possible to those of $p \left( \mathbf{x} \right)$.  To this
end, we need to choose proper values of $\mathbf{\theta}_i$ and
$\mathbf{\Phi}_i$ in order to achieve this objective.

The means and covariances of $p \left( \mathbf{x} \right)$ and
$\tilde{p} \left( \mathbf{x} \right)$, denoted by $\bar{\mathbf{x}}$
and $\bar{\mathbf{P}}$, and $\tilde{\mathbf{x}}$ and
$\tilde{\mathbf{P}}$, respectively, are given by
\begin{subequations} \label{eq:stat_original_pdf}
\begin{align}
\bar{\mathbf{x}} &= \sum_{i=1}^{N} w_i \mathbf{\mu}_i \, , ~\text{and}~
\bar{\mathbf{P}} = \sum_{s=1}^{N} w_i \left(\mathbf{\Sigma}_i + \left(  \mathbf{\mu}_i - \bar{\mathbf{x}} \right) \left(  \mathbf{\mu}_i - \bar{\mathbf{x}}  \right)^T \right) \, ,\\
\tilde{\mathbf{x}} &= \dfrac{1}{q} \sum_{i=1}^{q} \mathbf{\theta}_i \, , ~\text{and}~
\tilde{\mathbf{P}} = \dfrac{1}{q} \sum_{i=1}^{q} \left(\mathbf{\Phi}_i + \left(  \mathbf{\theta}_i - \tilde{\mathbf{x}} \right) \left(  \mathbf{\theta}_i - \tilde{\mathbf{x}}  \right)^T \right) \, .
\end{align}
\end{subequations}
Thus our objective is equivalent to balancing the above equation such that
\begin{equation} \label{eq:auxiliary_objective}
\tilde{\mathbf{x}} = \bar{\mathbf{x}} \, , ~\text{and}~ \tilde{\mathbf{P}} \approx \bar{\mathbf{P}} \, .
\end{equation}
In the trivial case with $q = N = 1$,
Eq.~(\ref{eq:auxiliary_objective}) can be satisfied by letting
$\mathbf{\theta}_1 = \mathbf{\mu}_1$ and $\mathbf{\Phi}_1 =
\mathbf{\Sigma}_1$, and the PEnKF reduces to an EnKF. In non-trivial
cases, for simplicity in solving Eq.~(\ref{eq:auxiliary_objective})
and reducing computational cost (as to be shown later), one may choose
the covariances $\mathbf{\Phi}_i$ to be constant, say
$\mathbf{\Phi}_i= \mathbf{\Phi}$, for $i=1,\dotsb,q$, so that
\begin{equation} \label{eq:objective_expression}
\dfrac{1}{q} \sum_{i=1}^{q} \mathbf{\theta}_i = \bar{\mathbf{x}} \, , ~\text{and}~
\mathbf{\Phi} + \dfrac{1}{q} \sum_{i=1}^{q} \left(  \mathbf{\theta}_i - \bar{\mathbf{x}} \right) \left(  \mathbf{\theta}_i - \bar{\mathbf{x}}  \right)^T \approx \bar{\mathbf{P}} \, .
\end{equation}

When an EnKF is used to construct the PKF, one needs to represent the
solution of Eq.~(\ref{eq:objective_expression}) in terms of some
ensembles $\{ \mathbf{X}_{en}^i, i=1,\dotsb,q\}$, where
$\mathbf{X}_{en}^i$ is a matrix containing the (analysis) ensemble of
the $i$th Gaussian component in Eq.~(\ref{eq:new_GMM}), with mean
$\mathbf{\theta}_i$ and covariance $\mathbf{\Phi}$. For simplicity, we
assume that $\mathbf{X}_{en}^i$ are all of dimension $n \times m$,
with the ensemble size $m$ for each $i$. Similar results can be easily
obtained in the case with non-uniform ensemble sizes.

We then define a constant $c$, called \emph{fraction coefficient}
hereafter, which satisfies that $0 \leq c \leq 1$. We let
$\mathbf{\Phi} \approx c^2 \bar{\mathbf{P}}$, so that
Eq.~(\ref{eq:objective_expression}) is reduced to
\begin{equation} \label{eq:reduced_objective_expression}
\dfrac{1}{q} \sum_{i=1}^{q} \mathbf{\theta}_i = \bar{\mathbf{x}} \, , ~\text{and}~
\dfrac{1}{q} \sum_{i=1}^{q} \left(  \mathbf{\theta}_i - \bar{\mathbf{x}} \right) \left(  \mathbf{\theta}_i - \bar{\mathbf{x}}  \right)^T \approx (1-c^2) \bar{\mathbf{P}} \, .
\end{equation}
In other words, the centers $\{ \mathbf{\theta}_i, i = 1, \dotsb, q
\}$ can be generated as a set of state vectors whose sample mean and
covariance are $\bar{\mathbf{x}}$ and $(1-c^2) \bar{\mathbf{P}}$,
respectively. After obtaining $\mathbf{\theta}_i$, one can generate
the corresponding ensembles $\mathbf{X}_{en}^i$, with the sample means
and covariances being $\mathbf{\theta}_i$ and $\mathbf{\Phi} \approx
c^2 \bar{\mathbf{P}}$, respectively. How $ \mathbf{\theta}_i$ and
$\mathbf{X}_{en}^i$ can be generated is discussed with more details in
the support material.

From the above discussion, we see that $c$ is a coefficient that
decides how to divide $\bar{\mathbf{P}}$ among $\mathbf{\Phi}$ and
$\dfrac{1}{q} \sum_{i=1}^{q} \left( \mathbf{\theta}_i -
  \bar{\mathbf{x}} \right) \left( \mathbf{\theta}_i - \bar{\mathbf{x}}
\right)^T$, so that the constraints in
Eq.~(\ref{eq:objective_expression}) are satisfied. When $c \rightarrow
0$, we have $\mathbf{\Phi} \rightarrow \mathbf{0}$ so that $\tilde{p}
\left( \mathbf{x} \right)$ in Eq.~(\ref{eq:new_GMM}) approaches the
Monte Carlo approximation in the particle filter, with the mass points
equal to $\mathbf{\theta}_i$. On the other hand, when $c \rightarrow
1$, we have $\dfrac{1}{q} \sum_{i=1}^{q} \left( \mathbf{\theta}_i -
  \bar{\mathbf{x}} \right) \left( \mathbf{\theta}_i - \bar{\mathbf{x}}
\right)^T \rightarrow \mathbf{0}$, so that all $\mathbf{\theta}_i$
approach $\bar{\mathbf{x}}$ and $\mathbf{\Phi}$ approaches
$\bar{\mathbf{P}}$. As a result, $\tilde{p} \left( \mathbf{x} \right)$
in Eq.~(\ref{eq:new_GMM}) approaches the Gaussian {\it pdf}
$N(\mathbf{x}: \bar{\mathbf{x}},\bar{\mathbf{P}})$, which is
essentially the assumption used in the EnKF. In this sense, when
equipped with the re-sampling algorithm, the PEnKF is a filter in
between the particle filter and the EnKF, with an adjustable parameter
$c$ that influences its behavior.

We note that, when $c \rightarrow 0$, under the constraint of matching the first two moments, our re-sampling scheme is very close to the posterior Gaussian re-sampling strategy used in the Gaussian particle filter \citep{Kotecha-signal-2003,Xiong-note-2006}, in which one generates particles from a Gaussian distribution with mean and covariance equal to those of the posterior {\it pdf} of the system states. As a result, there is no guarantee that higher order moments of the new MON match those of the original MON in our re-sampling scheme. If matching higher-order moments is a concern, one may adopt alternative criteria, for instance, the one that aims to minimize the distance (in certain metric) between the new MON and the original one, so that the re-sampling procedure is recast as an optimization problem, in which one aims to choose appropriate parameters, i.e., means and covariances of the new MON, that satisfy the chosen criterion as far as possible. In principle, this type of parameter estimation problem may be solved by the expectation-maximization (EM) algorithm \citep{Redner-mixture-1984,Smith-cluster}. But in practice, it is often computationally very intensive in doing so, due to the slow convergence rate of the EM algorithm and the high dimensionality of the parameter space in constructing the new MON. Therefore we do not consider this type of more sophisticated re-sampling strategy in this study.

For the purpose of pdf re-approximation, it is clear that the MON is not the only choice. A few alternatives are developed in the context of kernel density estimation (KDE) \citep{Silverman1986}, and in principle all of them can be applied for pdf re-approximation. For instance, KDE is adopted at the re-sampling step in the regularized
particle filter (RPF) \citep{Musso-improving-2001,Stavropoulos-improved-2001} to construct a continuous pdf with respect the particles before re-sampling, and to draw a number of new particles from the continuous pdf afterwards. In this regard, the PEnKF is similar to the RPF, especially if the Gaussian kernel is adopted in the RPF for density estimation. However, there also exist differences. We list some of them as follows. 
\begin{itemize}
\item The RPF first constructs a continuous pdf, and then draws a number of new particles with equal weights from the resulting pdf. In contrast, the PEnKF aims to directly approximate a MON by a new MON with equal weights.

\item In the RPF, various kernels can be adopted for the purpose of constructing the continuous pdf. However, in the PEnKF, we are confined to use the MON, since we aim to build the PEnKF consisting of a set of parallel EnKFs. 

\item The pdf re-approximation criterion used in the PEnKF only captures the first two moments of the underlying pdf. In contrast, KDE used in the RPF in principle can yield a very good pdf estimate, provided that there are sufficient particles. In certain circumstances, though, the number of required particles may also suffer from the ``curse-of-dimensionality'' \citep[ch.~4]{Silverman1986}.     

\end{itemize}

\subsection{Outline of the PEnKF Algorithm}

To facilitate the comprehension of the PEnKF, here we provide an
outline of the main steps of its algorithm. To avoid distraction, we
will discuss the initialization of the PEnKF in the next
section. Throughout this paper, we assume that the number $q$ of
Gaussian components at the re-sampling step and the number $N$ of
Gaussian components at the prediction and correction steps are time
invariant.  This implies the choice $q = N$. %If $q \neq N$ at the
%beginning, they will become equal to each other after the first
%re-sampling step.

Without loss of generality, we also assume that at time instant $k-1$,
the posterior pdf $p^a_{k-1} (\mathbf{x}_{k-1} \mid \yb_{1:k-1})$ is
re-approximated, through the re-sampling step, by a mixture model
\[
\tilde{p}_{k-1} (\mathbf{x}_{k-1} \mid \yb_{1:k-1}) =
\sum\limits_{i=1}^{q} \tilde{w}_{k-1}^i N \left(\mathbf{x}_{k-1}:
  \mathbf{\theta}_{k-1,i}, \mathbf{\Phi}_{k-1} \right) \, .
\]
Moreover, the re-approximated analysis ensembles $\{
\mathbf{X}_{approx}^{k-1,i}, i=1,\dotsb,q\}$ representing the Gaussian
components $N \left(\mathbf{x}_{k-1}: \mathbf{\theta}_{k-1,i},
  \mathbf{\Phi}_{k-1} \right)$ are also generated. The procedures at
the next assimilation cycle are outlined as follows.

\begin{enumerate}

\item[$\bullet$] {\underline {\it Prediction step}:} For
  $i=1,\dotsb,q$, propagate the ensembles
  $\mathbf{X}_{approx}^{k-1,i}$ forward through Eq.~(\ref{model}) to
  obtain the corresponding background ensembles
  $\mathbf{X}_{bg}^{k,i}$ at instant $k$. Accordingly, the background
  {\it pdf} becomes
\[
p^b_{k} (\mathbf{x}_k \mid \yb_{1:k-1}) = \sum\limits_{i=1}^{q}
\tilde{w}_{k-1}^i N \left(\mathbf{x}_k: \hat{\mathbf{x}}_{k,i}^b,
  \hat{\mathbf{P}}_{k,i}^b \right) \, ,
\]
with $\hat{\mathbf{x}}_{k,i}^b$ and $\hat{\mathbf{P}}_{k,i}^b$ being
the sample mean and covariance of the ensemble
$\mathbf{X}_{bg}^{k,i}$, respectively.

\item[$\bullet$] {\underline {\it Correction step}:} With an incoming
  observation $\yb_k$, for each background ensemble
  $\mathbf{X}_{bg}^{k,i}$, $i=1,\dotsb,q$, apply an EnKF to obtain the
  analysis mean $\hat{\mathbf{x}}_{k,i}^a$ and the analysis ensemble
  $\mathbf{X}_{ana}^{k,i}$. During the correction, covariance
  inflation and localization (cf. \S~\ref{sec:inflation and
    localization}) can be conducted on the EnKF. In addition, update
  the associated weights $\tilde{w}_{k-1}^i$ to $w_k^i$ according to
  Eq~(\ref{corrp}). After the corrections, the analysis {\it pdf}
  becomes
\[
p^a_{k} (\mathbf{x}_k \mid \yb_{1:k}) = \sum\limits_{i=1}^{q} w_k^i N
\left(\mathbf{x}_k: \hat{\mathbf{x}}_{k,i}^a, \hat{\mathbf{P}}_{k,i}^a
\right) \, ,
\]
where $w_k^i$ are computed according to Eq.~(\ref{corrp}) in the
context of the EnKF, and $\hat{\mathbf{P}}_{k,i}^a$ are the sample
covariances of $\mathbf{X}_{ana}^{k,i}$.

\item[$\bullet$] {\underline {\it Re-sampling step}:} Use the
  criterion in (\ref{eq:resampling_criterion}) to determine whether to
  conduct re-sampling or not.

\begin{enumerate}

\item[$(1)$] If there is no need for re-sampling, then assign
  $\tilde{p}_{k} (\mathbf{x}_k \mid \yb_{1:k}) = p^a_{k} (\mathbf{x}_k
  \mid \yb_{1:k})$, and $\mathbf{X}_{approx}^{k,i} =
  \mathbf{X}_{ana}^{k,i}$ for $i = 1, \dotsb, q$;

\item[$(2)$] Otherwise, $\tilde{p}_{k} (\mathbf{x}_k \mid \yb_{1:k}) =
  \dfrac{1}{q} \sum\limits_{i=1}^{q} N \left(\mathbf{x}_k:
    \mathbf{\theta}_{k,i}, \mathbf{\Phi}_{k} \right)$, where
  parameters $\mathbf{\theta}_{k,i}$ and $\mathbf{\Phi}_{k}$ are
  computed following the method in \S~\ref{sec:re-sampling}, and the
  associated weights become $1/q$. The ensembles
  $\mathbf{X}_{approx}^{k,i}$ are produced accordingly.

\end{enumerate}

\end{enumerate}

%##############################
%  NUMERICAL APPLICATION
%##############################

\section{Numerical Experiments}
\label{experiments}

\subsection{Experiment Design}
\label{exp:design}

In the present work, we focus on two different implementations of the
PEnKF: the first is based on the stochastic EnKF (SEnKF) of
\cite{Evensen-sequential} and the second based on the ensemble
transform Kalman filter (ETKF) of \cite{Bishop-adaptive}. These two
implementations are referred to as the PSEnKF and the PETKF,
respectively.

The strongly nonlinear $40$-dimensional system model due to
\cite{Lorenz-optimal} (LE98 model hereafter) was chosen as the testbed
to evaluate and study the performance of these two filters. This model
mimics the time-evolution of a scalar atmospheric quantity. It is
governed by the following set of equations:
\begin{equation} \label{LE98}
\frac{dx_i}{dt} = \left( x_{i+1} -
    x_{i-2} \right) x_{i-1} - x_i + 8, \, i=1, \dotsb, 40,
\end{equation}
where the nonlinear quadratic terms simulate advection and the linear
term represents dissipation. Boundary conditions are cyclic, i.e. we
define $x_{-1}=x_{39}$, $x_{0}=x_{40}$, and $x_{41}=x_{1}$. The model
was numerically integrated using the Runge-Kutta fourth order scheme
from time $t = 0$ to $t = 35$ with a constant time step $\Delta t =
0.05$ (which corresponds to $6$ hours in real time). To eliminate the
impact of transition states, the model trajectory between times $t =
0$ and $t = 25$ was discarded. The assimilation experiments were
carried out during the period $t = 25.05$ to $t = 35$ where the model
trajectory was considered to be the 'truth'.  Reference states were
then sampled from the true trajectory and a filter performance is
evaluated by how well it is able to estimate the reference states
using a perturbed model and assimilating a set of (perturbed)
observations that was extracted from the reference states.

In this work we consider two scenarios: one with a linear observation
operator and the other with a nonlinear operator. The concrete forms
of these two observational operators will be given in the relevant
sections below.

The time-averaged root mean squared error (rmse for short) is used to
evaluate the performance of a filter. Given a set of $n$-dimensional
state vectors $\{\mathbf{x}_k: \mathbf{x}_k =
(x_{k,1},\dotsb,x_{k,n})^T, k=0,\dotsb, k_{max} \}$, with $k_{max}$
being the maximum time index ($k_{max}=199$ in our experiments), then
the rmse $\hat{e}$ is defined as
\begin{equation}
  \hat{e} = \dfrac{1}{k_{max}+1} \sum\limits_{k=0}^{k_{max}} \sqrt{\dfrac{1}{n} \sum\limits_{i=1}^{n}
    (\hat{x}_{k,i}^a - x_{k,i})^2 }\, ,
\end{equation}
where $\hat{\mathbf{x}}_k^a =
(\hat{x}_{k,1}^a,\dotsb,\hat{x}_{k,n}^a)^T$ is the analysis state of
$\mathbf{x}_k$.

A possible problem in directly using $\hat{e}$ as the performance
measure is that $\hat{e}$ itself may depend on some intrinsic
parameters of the filters, for instance, the covariance inflation
factor and localization length scale as to be discussed later. This
may lead to inconsistent conclusions at different parameter values. To
avoid this problem, we adopted the following strategy: we relate a
filter's best possible performance to the minimum rmse
$\hat{e}_{min}$, which is the minimum value of $\hat{e}$ that the
filter can achieve within the chosen ranges of the filter's intrinsic
parameters. In performance comparison, if the minimum rmse
$\hat{e}_{min}^A$ of filter $A$ is less than the minimum rmse
$\hat{e}_{min}^B$ of filter $B$, filter $A$ is said to perform better
than filter $B$.

\subsection{Implementation Details}
\label{exp:ini}

\subsubsection{Filter Initialization}

To initialize the PEnKF, we first estimate the mean and covariance of
the LE98 model over some time interval following
\citet{Hoteit2008}. These statistics are then used to produce the {\it
  pdf} $p_0^f(\mathbf{x}_{0})$ of the background at the first
assimilation cycle as a MON.

Concretely, the LE98 model was first integrated for a long period
(between $t=0$ and $t=1000$) starting from an initial state that has
been drawn at random. The trajectory that falls between $t=50.05$ and
$t=1000$ was used to estimate the mean $\hat{\mathbf{x}}_{ds}$ and
covariance $\hat{\mathbf{P}}_{ds}$ of the dynamical system. To
initialize $p_0^f(\mathbf{x}_{0})$ as a mixture of $N$ Gaussian
distributions
\begin{equation}
  p_0^f(\mathbf{x}_{0}) = \dfrac{1}{N} \sum\limits_{i=1}^{N} N(\mathbf{x}_{0}:\mathbf{x}_{0}^{f,i} \, , \mathbf{P}_{com}) \, ,
\end{equation}
where $\mathbf{x}_{0}^{f,i}$ are the means, and $\mathbf{P}_{com}$ the
common covariance matrix of the Gaussian distributions in the mixture,
we draw $N$ samples $\mathbf{x}_{0}^{f,i}$ from the Gaussian
distribution $N(\mathbf{x}_{0}:\hat{\mathbf{x}}_{ds},
\hat{\mathbf{P}}_{ds})$, and set $\mathbf{P}_{com} =
\hat{\mathbf{P}}_{ds}$. If $\hat{\mathbf{x}}_{0}^{f} = \dfrac{1}{N}
\sum\limits_{i=1}^{N} \mathbf{x}_{0}^{f,i}$ denotes the sample mean of
$\mathbf{x}_{0}^{f,i}$, then the covariance $\mathbf{P}_{0}^f$ of
$p_0^f(\mathbf{x}_{0})$ is given by
\begin{equation}
  \mathbf{P}_{0}^f = \hat{\mathbf{P}}_{ds} + \dfrac{1}{N} \sum\limits_{i=1}^{N} (\mathbf{x}_{0}^{f,i} - \hat{\mathbf{x}}_{0}^{f} )(\mathbf{x}_{0}^{f,i} - \hat{\mathbf{x}}_{0}^{f} )^T \, ,
\end{equation}
which is always larger than $\hat{\mathbf{P}}_{ds}$. The rationale
behind this choice is not far from the covariance inflation technique
\citep{Anderson-Monte,Whitaker-ensemble}. In practice, a data
assimilation system is often subject to various errors, such as poorly
known model and observational errors, sampling errors, etc. In such
circumstances, an inflated background covariance would allocate more
weights to the incoming observation when updating the background to
the analysis, making the filter more robust
\citep{Jazwinski1970,Simon2006}.

\subsubsection{Covariance Inflation and Localization} \label{sec:inflation and localization}

Covariance inflation \citep{Anderson-Monte,Whitaker-ensemble} and
localization \citep{Hamill-distance} are two popular techniques that
are used to improve the stability and performance of the EnKF
\citep{Hamill2009,VanLeeuwen2009}, especially in the small ensemble
scenario. In our experiments, these two techniques are implemented for
each EnKF in the PEnKF.

More concretely, to introduce covariance inflation to the $i$th EnKF
at instant $k$, we multiply the analysis covariance
$\hat{\mathbf{P}}_{k,i}^a$ (before the re-sampling step) by a factor
$(1+\delta)^2$, where the scalar $\delta \geq 0$, called {\it
  covariance inflation factor}, is introduced as an intrinsic
parameter of the EnKF. On the other hand, we follow the method in
\citet{Hamill-distance} to conduct covariance localization on the
background covariance and its projection onto the observation space,
with the tapering function (for smoothing out spuriously large values
in covariance matrices) being the fifth order function defined in
Eq.~(4.10) of \citet{Gaspari1999}. In doing so, another intrinsic
scalar parameter $l_c > 0$, called {\it length scale}
\citep{Hamill-distance}, is introduced to the EnKF. Roughly speaking,
$l_c$ is a parameter that determines the critical distance beyond
which the tapering function becomes zero.

\subsection{Experiments Results with a Linear Observation Operator}
\label{exp_linear:results}

In the first scenario, we let the (synthetic) observations be generated every day ($4$ model time
steps) from the reference states using the following linear
observation system
\begin{equation} \label{sim:lin_observer}
\mathbf{y}_k = (x_{k,1},x_{k,3},\dotsb,x_{k,39})^T + \mathbf{v}_k\, ,
\end{equation}
where only the odd state variables $x_{k,i}$ ($i=1,3,\dotsb,39$) of
the system state $\mathbf{x}_k \equiv (x_{k,1},\dotsb,x_{k,40})^T$ at
time index $k$ are observed. The observation noise $\mathbf{v}_k$
follows the $20$-dimensional Gaussian distribution $N(\mathbf{v}_k:
\mathbf{0}, \mathbf{I}_{20})$ with $\mathbf{I}_{20}$ being the $20
\times 20$ identity matrix.

\subsubsection{Effect of the Number of Gaussian Distributions}
\label{sec:lin_exp_c_vs_q}

In the first experiment we examine the effect of the number of
Gaussian distributions on the performance of the PSEnKF and the
PETKF. The experiment settings are as follows.

We initialize the pdf $p_0^f(\mathbf{x}_{0})$ with $N$ Gaussian {\it
  pdf}s. In our experiments we let $N$ take values between $1$ and
$60$. Since it is costly to carry out the computation for each integer
in this interval, we choose to let $N$ increase from $1$ to $10$, with
an even increment of $1$ each time, and then increase it from $15$ to
$60$, with a larger increment of $5$ each time, as $N$ becomes
larger. For convenience, we denote this choice by $N \in
\{1:1:10,15:5:60\}$, where the notation $v_{min}:v_{inc}:v_{max}$
represents a set of values increasing from $v_{min}$ to $v_{max}$,
with an even increment of $v_{inc}$ each time. If there is a need to
conduct re-sampling, we re-approximate the analysis MON by a new MON
with equal weights and with the same number of normal distributions.
In doing so, we introduce a new parameter, i.e., the
fraction coefficient $c$ defined in \S~ \ref{sec:re-sampling}, to the
PSEnKF/PETKF. To examine its effect on the performance of the filter,
we let $c \in \{0.05:0.1:0.95\}$. The ensemble size is set to $m=20$
in each SEnKF/ETKF, which is relatively small compared to the system
dimension $40$. In this case, it is customary to conduct covariance
inflation \citep{Anderson-Monte,Whitaker-ensemble} and localization
\citep{Hamill-distance} to improve the robustness and performance of
the filters \citep{Hamill2009,VanLeeuwen2009}. The impacts of
covariance inflation and localization on the performance of the EnKF
have been examined in many works, see, for example,
\citet{Whitaker-ensemble}. In our experiments we let the covariance
inflation factor $\delta=0.02$. We follow the settings in
\citet[\S~7.2.3]{Luo2008-spgsf1} to conduct covariance localization
and choose the length scale $l_c=50$. To reduce statistical
fluctuations, we repeat the experiments for $20$ times, each time with
a randomly drawn initial background ensemble, but the same true
trajectory and the corresponding observations. The same repetition
setting is adopted in all the other experiments.

\begin{figure*} %-- 1 --
\vspace*{2mm}
\centering
%--
	\subfigure[Stochastic EnKF-based PKF]{
	\label{fig_lin:sEnKF_based_GMM_contour_fraction_vs_nGMM_grayColor}	
	\includegraphics[width=0.8\textwidth]
{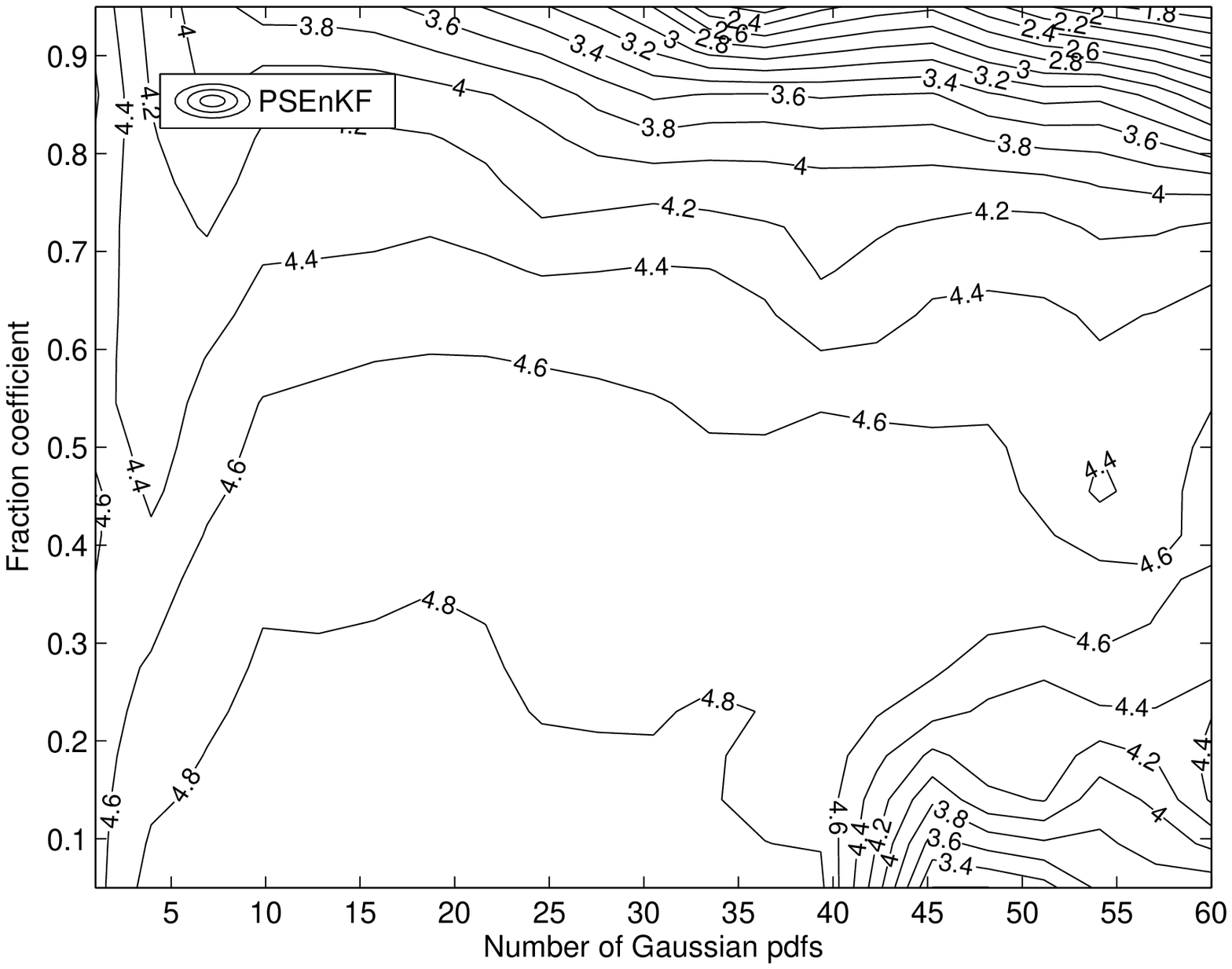}
     }
%--
	\subfigure[ETKF-based PKF]{
	\label{fig_lin:ETKF_based_GMM_contour_fraction_vs_nGMM_grayColor}	
	\includegraphics[width=0.8\textwidth]
{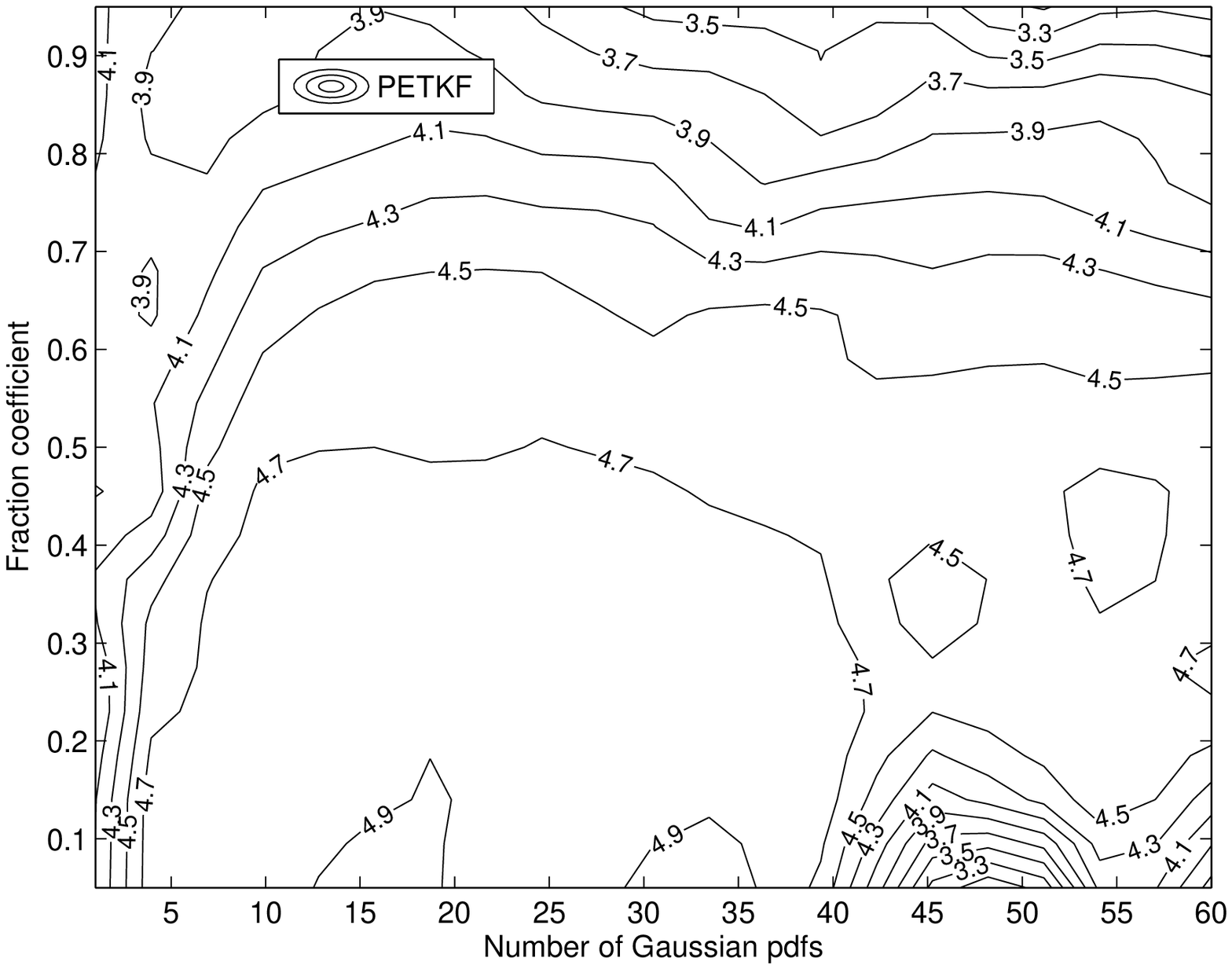}
     }
\caption{\label{fig_lin:GMM_contour_fraction_vs_nGMM_grayColor} RMS errors (over $20$ experiments) of the stochastic EnKF- and ETKF-based PEnKFs (with a fixed ensemble size of $20$ in each ensemble filter) as the functions of the fraction coefficient and the number of Gaussian {it pdf}s in the MON.}
\end{figure*}

In Fig.~\ref{fig_lin:GMM_contour_fraction_vs_nGMM_grayColor} we show
the rms errors of both the PSEnKF and PETKF as functions of the
fraction coefficient $c$ and the number $N$ of Gaussian {\it
  pdf}s. First, we examine how the rmse of the PSEnKF changes with $c$ when $N$ is fixed. In Fig.~\ref{fig_lin:sEnKF_based_GMM_contour_fraction_vs_nGMM_grayColor}, if $N$ is relatively small (say $N<40$), the rmse tends to decrease as $c$ increases. For larger $N$ (say $N=55$), the rmse of the filter exhibits the bell-shape behavior: at the beginning it increases when $c$ grows from $0$; after $c$ becomes relatively large (say $c=0.4$), further increasing $c$ reduces the rmse instead. Next, we examine the behavior of the rmse of the PSEnKF with respect to $N$ when $c$ is fixed. When $c$ is relatively small (say $c=0.1$), the rmse exhibits the U-turn behavior: at the beginning it intends to decrease as $N$ grows; after $N$ becomes relatively large (say $N=45$), further increasing $N$ increases the rmse instead. When $c$ is larger, say, $c=0.6$, the rmse appears less sensitive to the change of $N$. However, for even larger values of $c$, say, $c=0.9$, the rmse appears to monotonically decrease with $N$.

The behavior of the PETKF (cf. Fig.~\ref{fig_lin:ETKF_based_GMM_contour_fraction_vs_nGMM_grayColor}) with respect to the changes of $N$ and $c$ is similar to that of the PSEnKF. Therefore we do not repeat its description here.

\begin{figure*} %-- 2 --
\vspace*{2mm}
\centering

\includegraphics[width=0.8\textwidth]{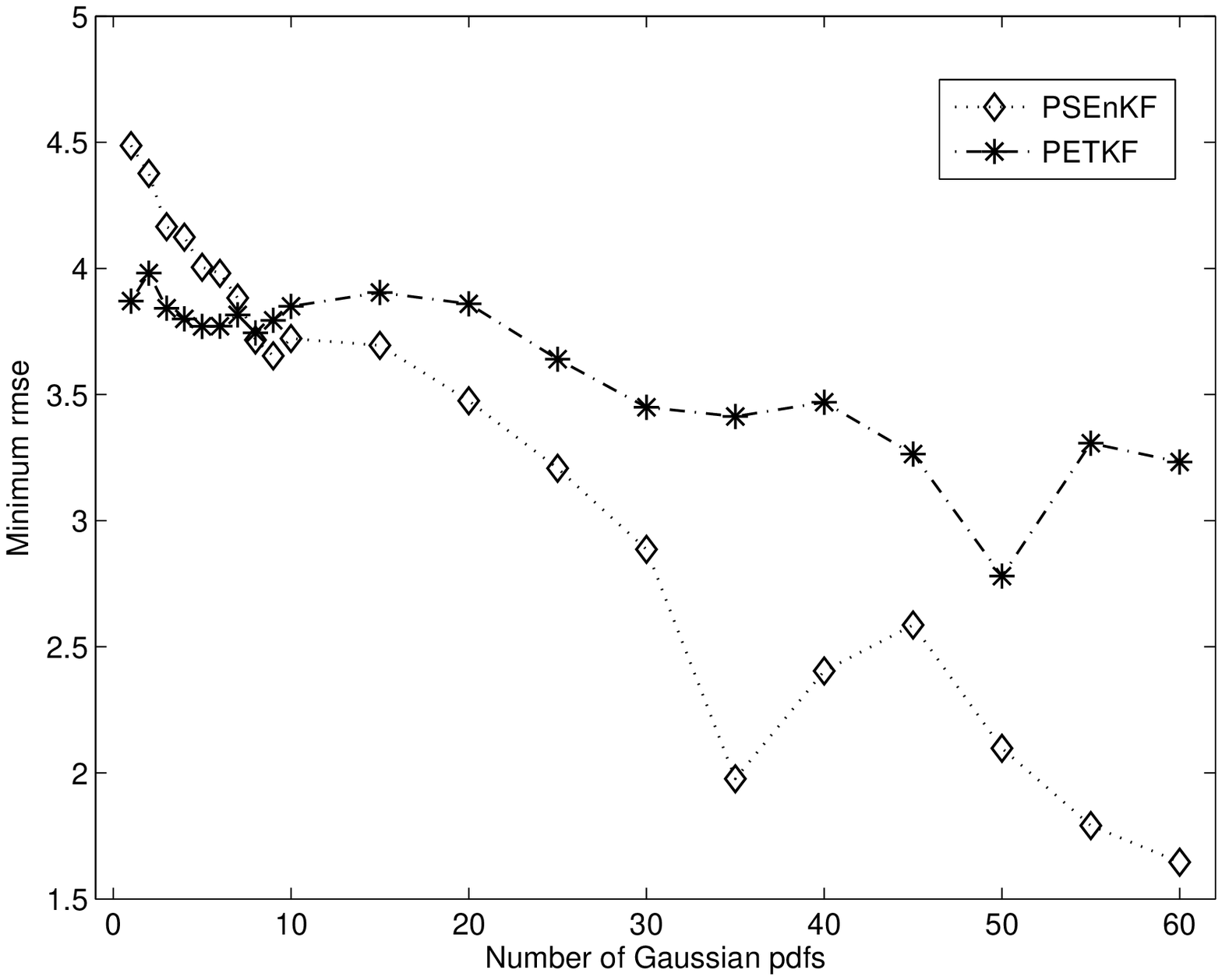}

\caption{\label{fig_lin:sEnKF_ETKF_based_GMM_minRMSE_vs_nGMM_grayColor}
  Minimum rms errors $\hat{e}_{min}$ (over $20$ experiments) of the
  stochastic EnKF- and ETKF-based PEnKFs (with a fixed ensemble size of
  $20$ in each ensemble filter) as the function of the number of
  Gaussian {\it pdf}s in the MON.}
\end{figure*}

To examine the minimum rms errors $\hat{e}_{min}$ of the PSEnKF and
the PETKF within the tested values of $c$ and $N$, we plot
$\hat{e}_{min}$ of both filters as functions of $N$ in
Fig.~\ref{fig_lin:sEnKF_ETKF_based_GMM_minRMSE_vs_nGMM_grayColor}. The
$\hat{e}_{min}$ of both filters tends to decrease as the number $N$ of
Gaussian distributions increases, though there also exhibit certain local minima. The PSEnKF achieves its lowest $\hat{e}_{min}$ at $N=60$, while the PETKF at $N=50$. As $N$ grows, both the PSEnKF
and the PETKF tend to have lower $\hat{e}_{min}$ than their corresponding base
filters, the SEnKF and the ETKF (corresponding to the PSEnKF and the
PETKF with $N=1$, as discussed in \S~\ref{sec:re-sampling}),
respectively. This confirms the benefit of accuracy improvement by
using the PEnKF instead of an EnKF. A comparison between the PSEnKF
and the PETKF shows that the PETKF performs better than the PSEnKF
when the number $N$ of Gaussian distributions is relatively small
(say, $N \leq 7$). However, as $N$ becomes larger, the PSEnKF
outperforms its ETKF-based counterpart instead. Similar phenomena can
also be observed in other experiments, as to be shown later.

\subsubsection{Effect of the Ensemble Size} \label{sec:lin_exp_n_vs_c}

In the second experiment we examine the effect of the ensemble size of
each SEnKF/ETKF in the PEnKF, on the performance of the
PSEnKF/PETKF. For reference, we also examine the performance of the
SEnKF and the ETKF under various ensemble sizes. The experiment
settings are as follows. For the PSEnKF and the PETKF, we let the
ensemble size $m$ of each EnKF take values from the set
$\{20,40,80,100,200,400,800,1000\}$. For a single SEnKF/ETKF, we
let $m \in \{20,40,60,80,100,200,400,600,800,1000\}$, with two more values at $60$ and $600$.

In the PSEnKF and the PETKF, we also vary the fraction coefficient $c$
such that $c \in \{0.05:0.1:0.95\}$. We fix the number $N$ of Gaussian
{\it pdf}s, i.e., the number of ensemble filters, to be $3$. To
conduct covariance inflation, we let the inflation factor
$\delta=0.02$. We choose to conduct covariance localization, and set
the length scale $l_c=50$, only if the ensemble size $m$ is not larger
than the dimension $40$ of the LE98 model. No covariance localization
was conducted if $m>40$. Our experience shows that, for $m>40$, the
benefit of conducting localization is not significant even if the
length scale $l_c$ is properly chosen, while an improper value of
$l_c$ is more likely to deteriorate the filter performance. To reduce
statistical fluctuations, the experiments are again repeated for $20$
times.

\begin{figure*} %-- 3 --
\vspace*{2mm}
\centering

\includegraphics[width=0.8\textwidth]{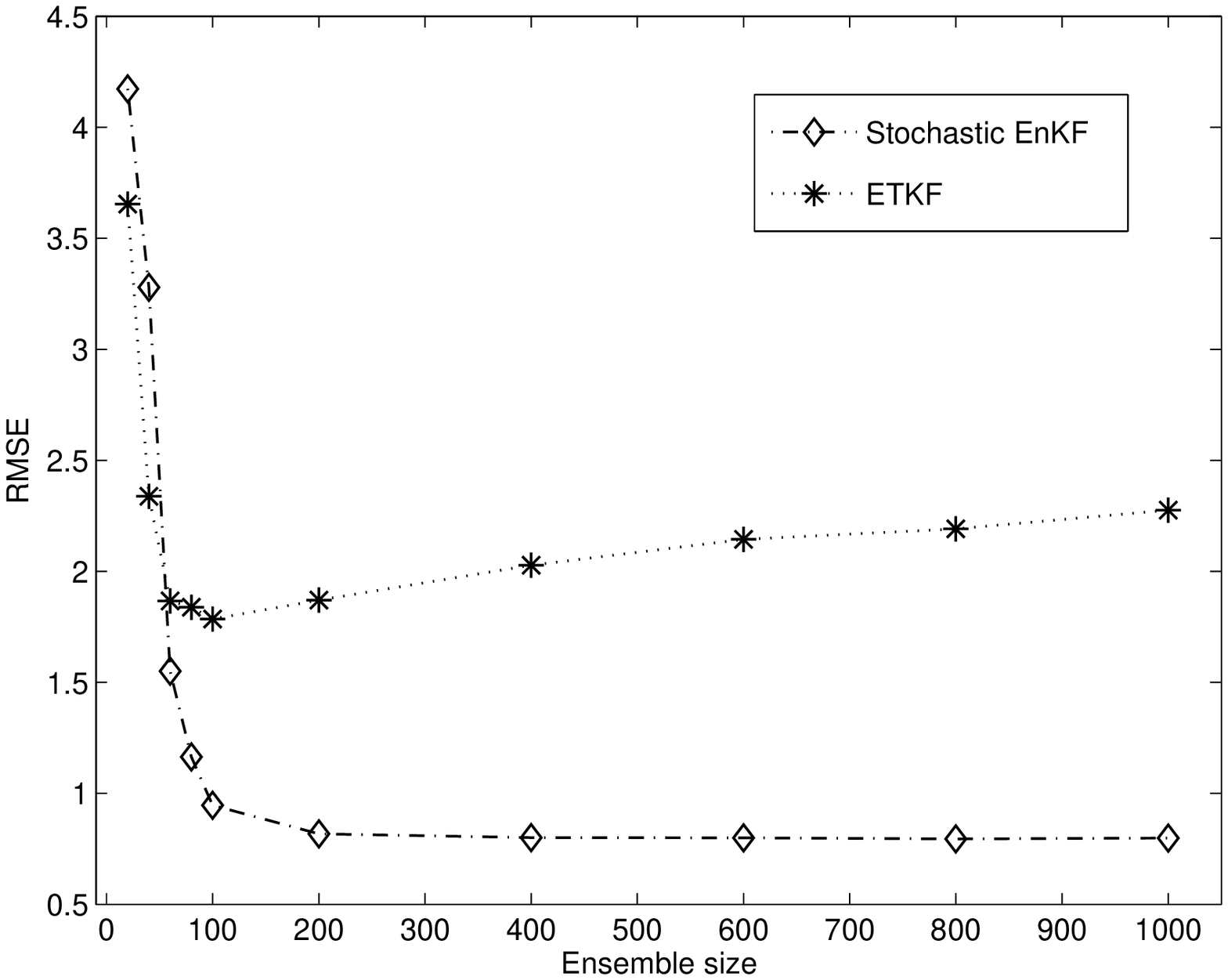}

\caption{\label{fig_lin:sEnKF_ETKF_RMSE_vs_nEn_grayColor} RMS errors (over
  $20$ experiments) of the stochastic EnKF and the ETKF as the
  functions of the ensemble size.}
\end{figure*}

In Fig.~\ref{fig_lin:sEnKF_ETKF_RMSE_vs_nEn_grayColor} we show the rms
errors of the SEnKF and the ETKF as functions of the ensemble size
$m$. The rmse of the ETKF exhibits a U-turn behavior. The rmse of the
ETKF monotonically decreases as long as $m < 100$. Beyond that, the
rmse monotonically increases instead as $m$ increases. On the other
hand, the SEnKF exhibits a different behavior. Its rmse decreases for
$m \leq 200$, and then reaches a plateau where the rmse remains almost
unchanged as $m$ further increases.

\begin{figure*} %-- 4 --
\vspace*{2mm}
\centering
%--
	\subfigure[Stochastic EnKF-based PKF]{
	\label{lin_sEnKF_based_GMM_contour_fraction_vs_nEn_grayColor}	
	\includegraphics[width=0.8\textwidth]
{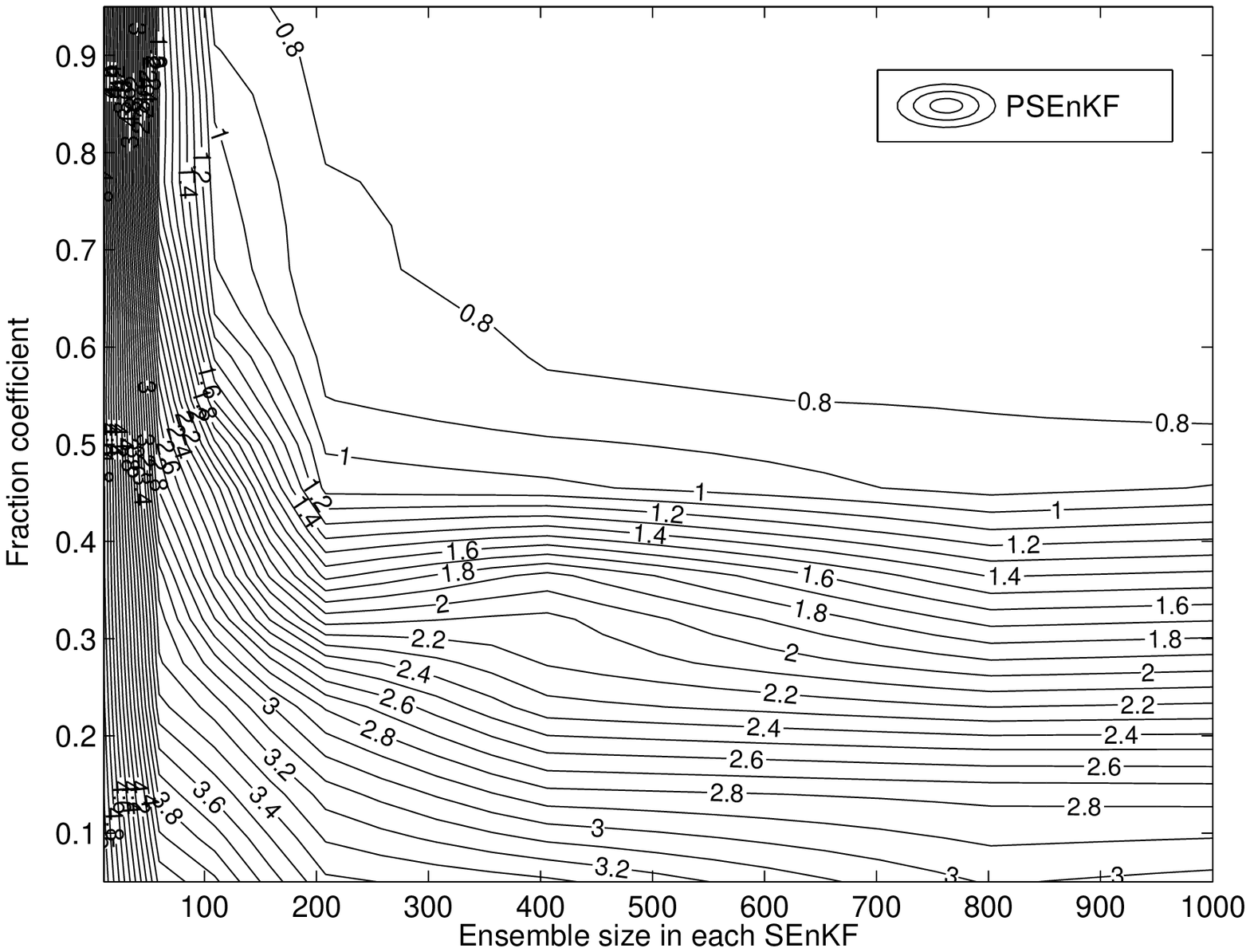}
     }
%--
	\subfigure[ETKF-based PKF]{
	\label{fig_lin:ETKF_based_GMM_contour_fraction_vs_nEn_grayColor}	
	\includegraphics[width=0.8\textwidth]
{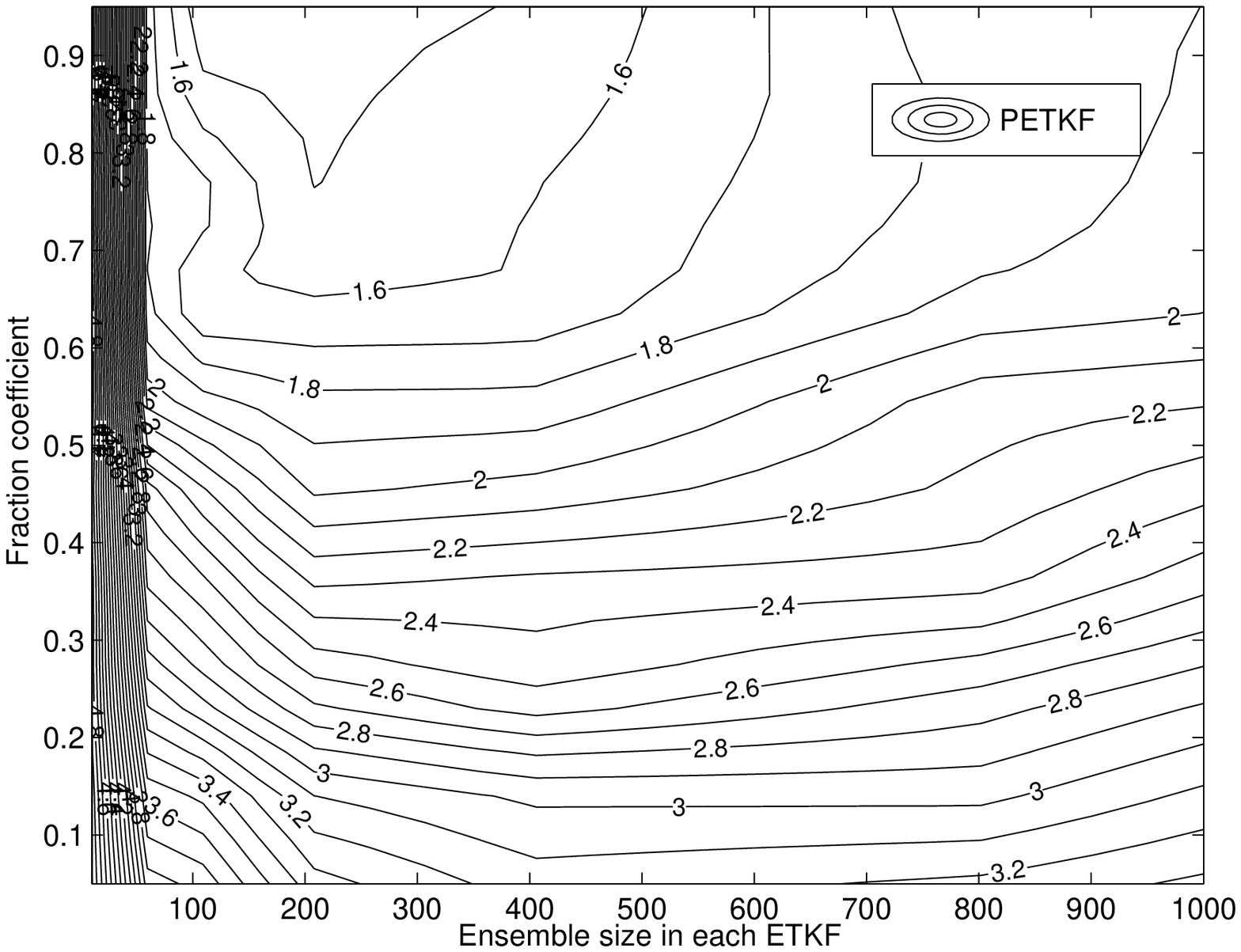}
     }
     \caption{\label{fig_lin:GMM_contour_fraction_vs_nEn_grayColor} RMS
       errors (over $20$ experiments) of the stochastic EnKF- and
       ETKF-based PEnKFs (with a fixed number of Gaussian {\it pdf}s of $3$ in
       each PKF) as the functions of the fraction coefficient and the
       ensemble size of the ensemble filter.}
\end{figure*}

Fig.~\ref{fig_lin:GMM_contour_fraction_vs_nEn_grayColor} plots the rms
errors of the PSEnKF and the PETKF as functions of the fraction
coefficient $c$, and the ensemble size $m$ in the SEnKF and the ETKF
used to construct the corresponding PEnKFs. The rms errors, as
functions of the ensemble size $m$ (with fixed $c$), are consistent
with our observations in
Fig.~\ref{fig_lin:sEnKF_ETKF_RMSE_vs_nEn_grayColor}. On the other
hand, for both PEnKFs, their rms errors tend to decrease as the
fraction coefficient $c$ increases.

\begin{figure*} % -- 5 --
\vspace*{2mm}
\centering

\includegraphics[width=0.8\textwidth]{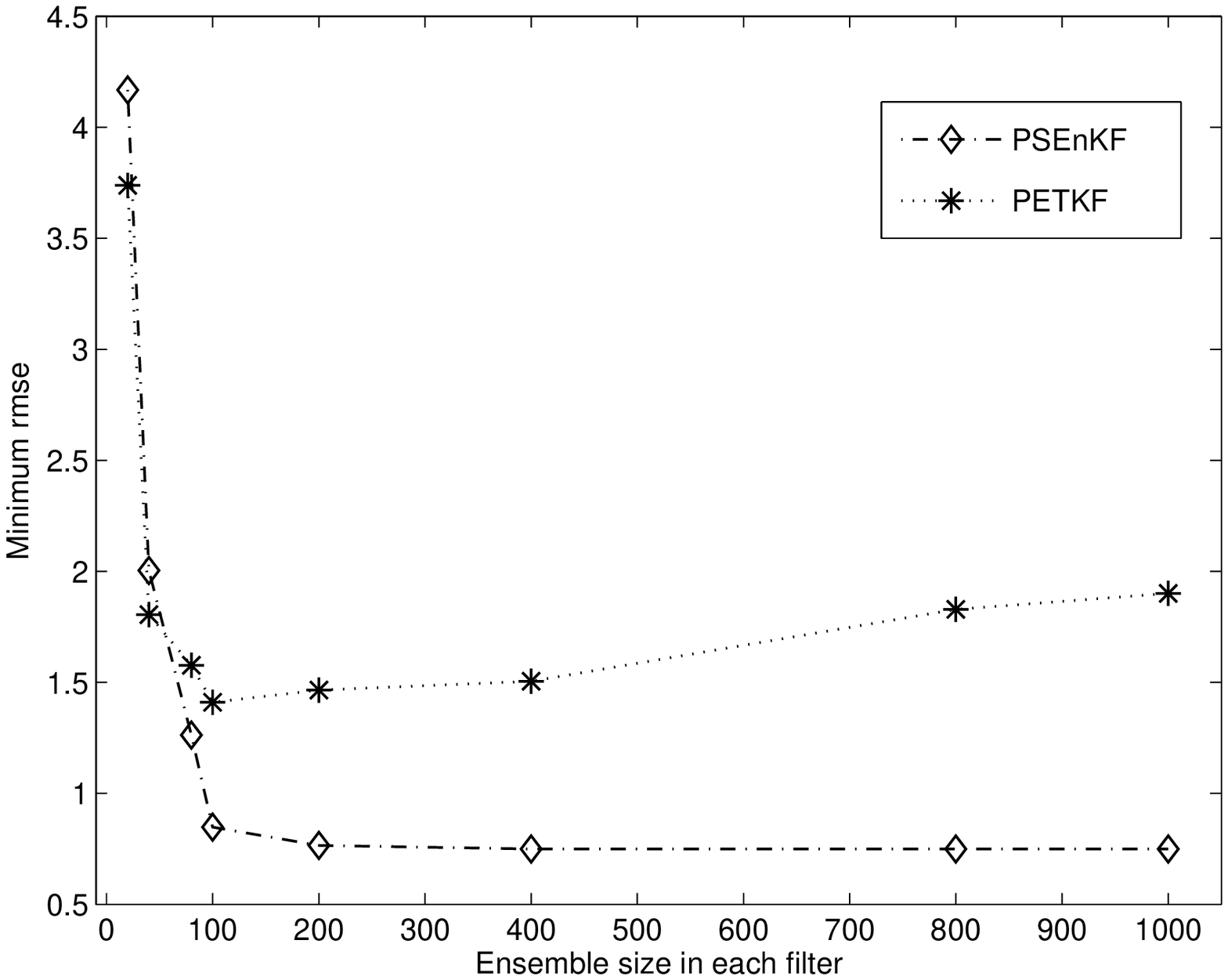}

\caption{\label{fig_lin:sEnKF_ETKF_based_GMM_minRMSE_vs_nEn_grayColor}
  Minimum rms errors $\hat{e}_{min}$ (over $20$ experiments) of the
  stochastic EnKF- and ETKF-based PEnKFs (with a fixed number of
  Gaussian {\it pdf}s of $3$ in each PKF) as the function of the ensemble
  size in each ensemble filter.}
\end{figure*}

Per analogy to the first experiment,
Fig.~\ref{fig_lin:sEnKF_ETKF_based_GMM_minRMSE_vs_nEn_grayColor} plots
the minimum rms errors $\hat{e}_{min}$ of the PSEnKF and the PETKF
within the tested fraction coefficient $c$ and the ensemble size
$m$. A comparison between
Figs.~\ref{fig_lin:sEnKF_ETKF_based_GMM_minRMSE_vs_nEn_grayColor} and
\ref{fig_lin:sEnKF_ETKF_RMSE_vs_nEn_grayColor} shows that, the minimum
rms errors $\hat{e}_{min}$ of the PEnKFs behave very similarly to the
rms errors of their corresponding EnKFs in
Fig.~\ref{fig_lin:sEnKF_ETKF_RMSE_vs_nEn_grayColor}. Moreover, the values of $\hat{e}_{min}$ in
Fig.~\ref{fig_lin:sEnKF_ETKF_based_GMM_minRMSE_vs_nEn_grayColor} tends to be
lower than the corresponding rms errors in
Fig.~\ref{fig_lin:sEnKF_ETKF_RMSE_vs_nEn_grayColor}, indicating the
benefit of accuracy improvement in using the PEnKFs. Again, a
comparison between the PSEnKF and the PETKF shows that the PETKF
performs better than the PSEnKF when the ensemble size $m$ is
relatively small (say, $m \leq 40$). However, as $m$ becomes larger,
the PSEnKF outperforms the PETKF instead.

%--nonlinear observers--
\subsection{Experiments Results with a Nonlinear Observation Operator}
\label{exp_nln:results}

In the second scenario, we introduce nonlinearity to the
observation system. To this end, we let the observations be generated
by the following nonlinear process
\begin{equation} \label{sim:nln_observer}
\mathbf{y}_k = 0.05(x_{k,1}^2,\dotsb,x_{k,39}^2)^T + \mathbf{v}_k\,
\end{equation}
for every $4$ model time steps. In Eq.~(\ref{sim:nln_observer}), again
only the odd state variables $x_{k,i}$ ($i=1,3,\dotsb,39$) of the
system state $\mathbf{x}_k \equiv (x_{k,1},\dotsb,x_{k,40})^T$ at time
index $k$ are observed. The observation noise $\mathbf{v}_k$ also
follows the $20$-dimensional Gaussian distribution
$N(\mathbf{v}_k:\mathbf{0}, \mathbf{I}_{20})$. We conduct the same
experiments as those in the case of linear observation operator.

\subsubsection{Effect of the Number of Gaussian Distributions}
\label{sec:nin_exp_c_vs_q}

We first examine the effect of the number of Gaussian
distributions. The experiment settings are the same as those in
\S~\ref{sec:lin_exp_c_vs_q}. Concretely, For either the PSEnKF or the
PETKF, the number of Gaussian distributions $N \in
\{1:1:10,15:5:60\}$, the fraction coefficient $c \in
\{0.05:0.1:0.95\}$. For each individual SEnKF/ETKF in the PEnKF, the
ensemble size $m=20$, the covariance inflation factor $\delta=0.02$
and the length scale $l_c=50$ for covariance localization. As before,
the experiments are repeated for $20$ times to reduce statistical
fluctuations.

\begin{figure*} %-- 6 --
\vspace*{2mm}
\centering
%--
	\subfigure[Stochastic EnKF-based PKF]{
	\label{fig:sEnKF_based_GMM_contour_fraction_vs_nGMM_grayColor}	
	\includegraphics[width=0.8\textwidth]
{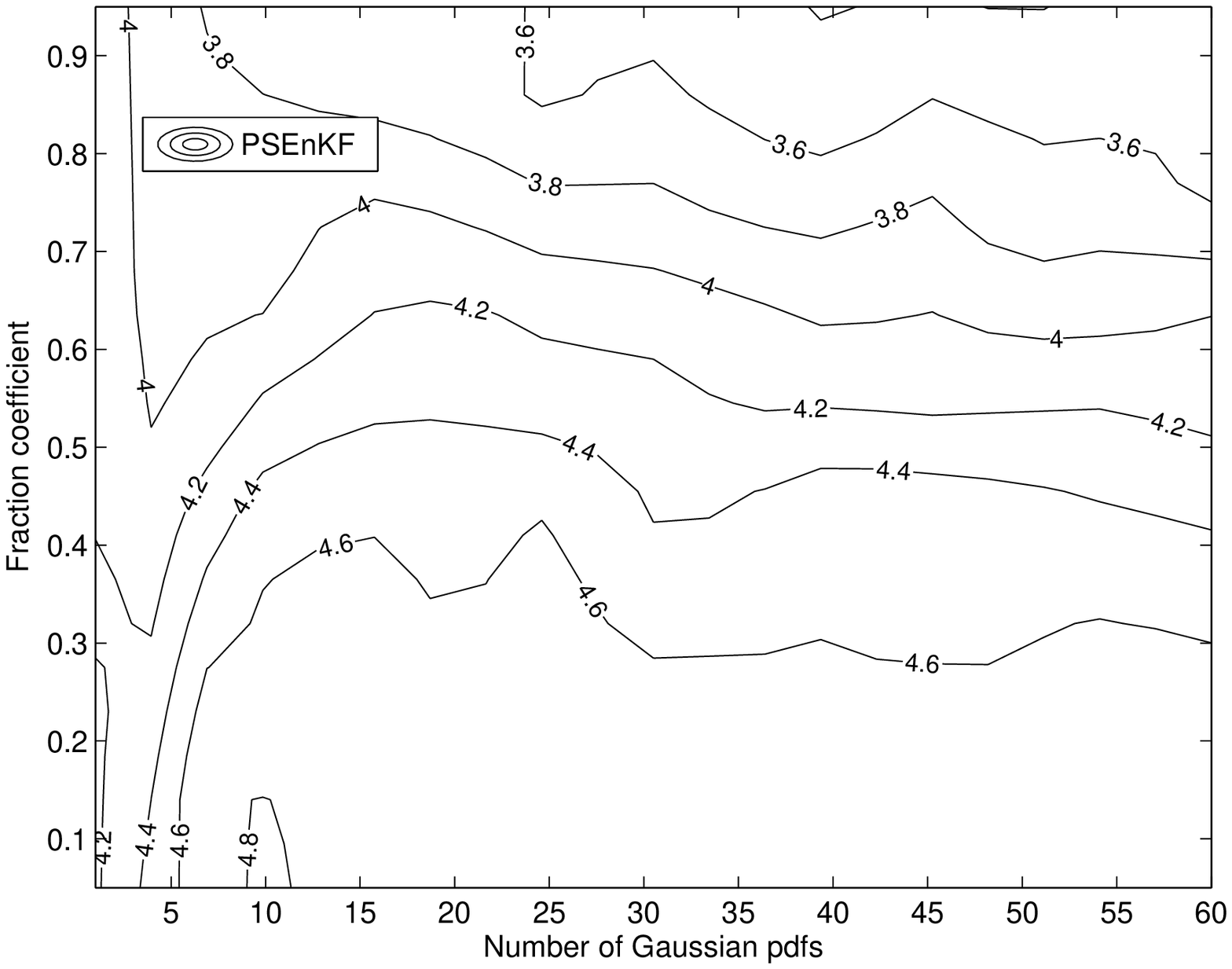}
     }
%--
	\subfigure[ETKF-based PKF]{
	\label{fig:ETKF_based_GMM_contour_fraction_vs_nGMM_grayColor}	
	\includegraphics[width=0.8\textwidth]
{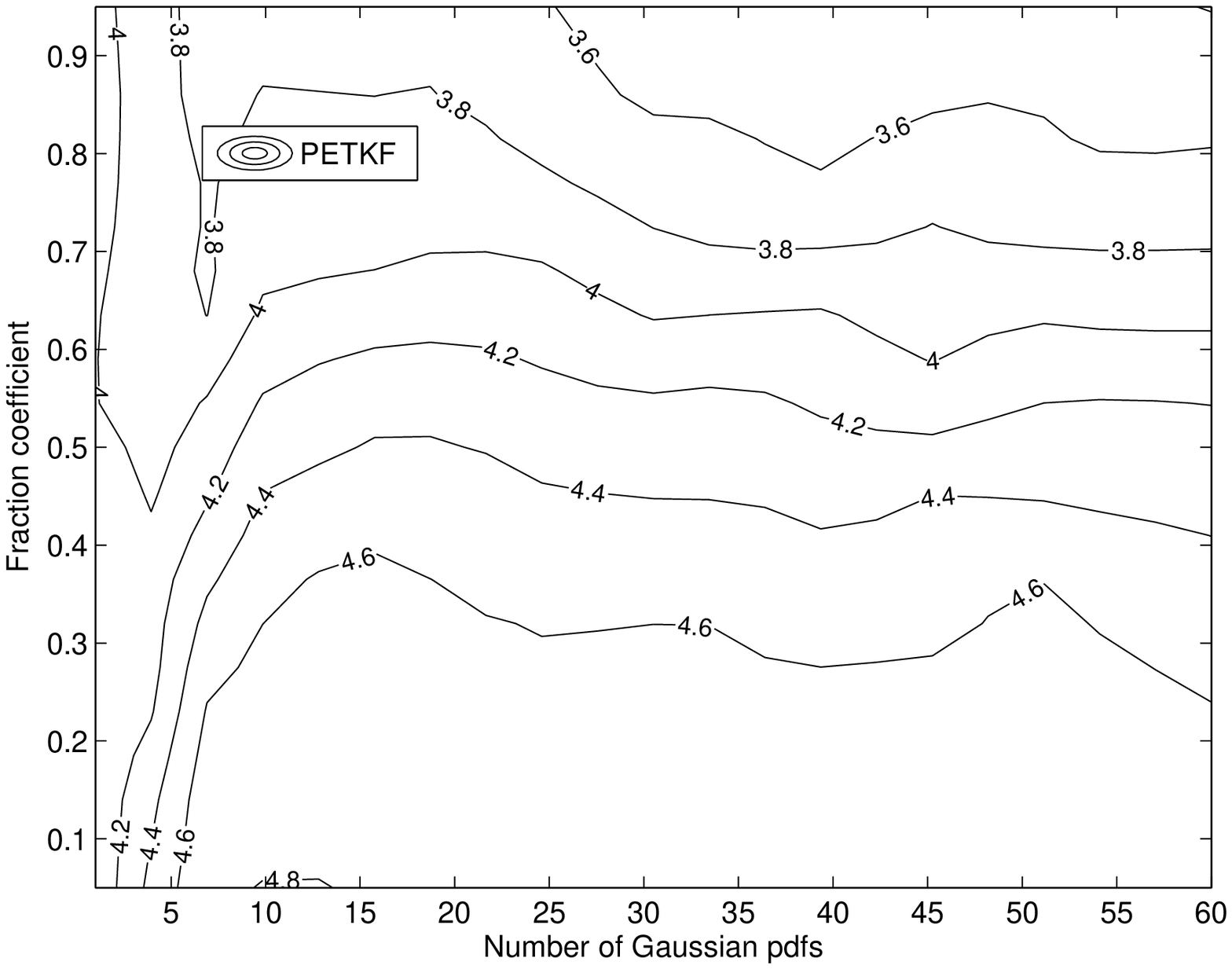}
     }
\caption{\label{fig:GMM_contour_fraction_vs_nGMM_grayColor} RMS errors (over $20$ experiments) of the stochastic EnKF- and ETKF-based PEnKFs (with a fixed ensemble size of $20$ in each ensemble filter) as the functions of the fraction coefficient and the number of Gaussian {\it pdf}s in the MON.}
\end{figure*}

Fig.~\ref{fig:GMM_contour_fraction_vs_nGMM_grayColor} plots the rms
errors of both the PSEnKF and the PETKF as functions of the fraction
coefficient $c$ and the number $N$ of Gaussian {\it pdf}s. When $c$ and $N$ changes, both the PSEnKF and the PETKF behave very similar to their counterparts in the linear case. The rms errors of the filters tend to decrease as $N$ increases, meaning that the PSEnKF/PETKF with $N>1$ in general performs better than the stochastic EnKF /ETKF
(corresponding to the case $N=1$ in the PEnKF), consistent with the results obtained in the linear observer case.

\begin{figure*} %-- 7 --
\vspace*{2mm}
\centering

\includegraphics[width=0.8\textwidth]{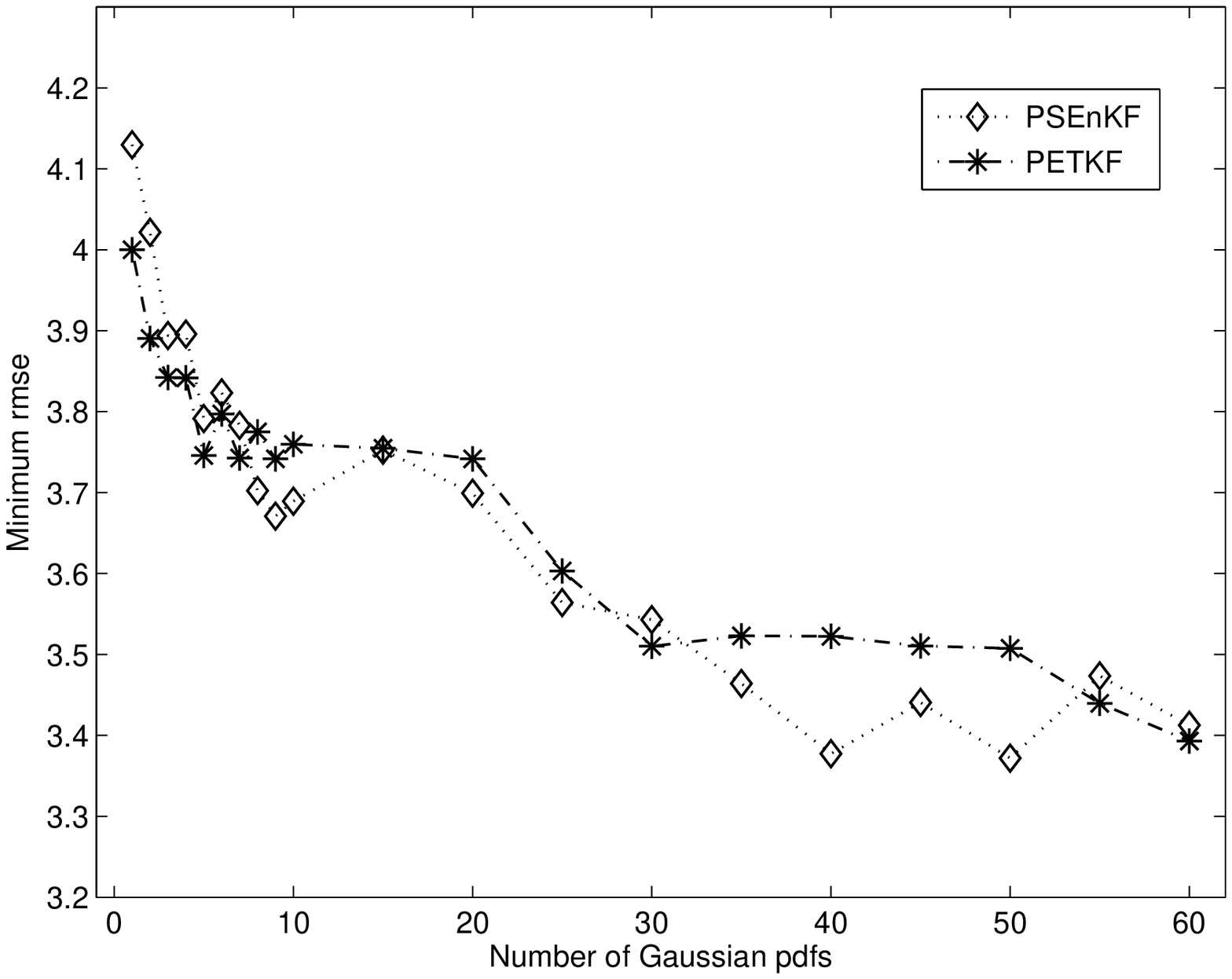}

\caption{\label{fig:sEnKF_ETKF_based_GMM_minRMSE_vs_nGMM_grayColor}
  Minimum rms errors $\hat{e}_{min}$ (over $20$ experiments) of the
  stochastic EnKF- and ETKF-based PEnKFs (with a fixed ensemble size of
  $20$ in each ensemble filter) as the function of the number of
  Gaussian {\it pdf}s in the MON.}
\end{figure*}

We also examine the minimum rms errors $\hat{e}_{min}$ of the PSEnKF
and the PETKF within the tested values of $c$ and
$N$. Fig.~\ref{fig:sEnKF_ETKF_based_GMM_minRMSE_vs_nGMM_grayColor}
plots $\hat{e}_{min}$ as functions of $N$. For the PSEnKF, the lowest
$\hat{e}_{min}$ is achieved at $N=50$.
And for the PETKF, its $\hat{e}_{min}$ tends to decrease within
the tested range of $N$, and achieves its minimum at $N=60$. The PEnKF with more than
one Gaussian distributions ($N>1$) performs better than the
corresponding EnKF ($N=1$). In addition, a comparison between the
PSEnKF and the PETKF shows again that the PETKF performs better than
the PSEnKF when the number $N$ of Gaussian distributions is relatively
small, but tends to become worse as $N$ increases.

A comparison between Figs.~\ref{fig_lin:sEnKF_ETKF_based_GMM_minRMSE_vs_nGMM_grayColor} and \ref{fig:sEnKF_ETKF_based_GMM_minRMSE_vs_nGMM_grayColor} shows that the rmse of a filter (e.g. the PSEnKF at $N=2$) with a nonlinear observer sometimes may be lower than that of the same filter with a linear observer \footnote{The result of comparison would also depend on the filter in use, its configuration, the system in assimilation, and so on, and therefore may change from case to case.}. This seemingly counter-intuitive result happens possibly because in such situations, the effect of sampling error due to the relatively small ensemble size dominates the effect of nonlinearity in the observation system. However, as the number $N$ of Gaussian distributions increases, the effect of nonlinearity becomes more prominent so that the rmse with a nonlinear observer tends to be higher than that with a linear one. Similar phenomenon can also be found by comparing Figs.~\ref{fig_lin:sEnKF_ETKF_RMSE_vs_nEn_grayColor} and \ref{fig_lin:sEnKF_ETKF_based_GMM_minRMSE_vs_nEn_grayColor} with Figs.~\ref{fig:sEnKF_ETKF_RMSE_vs_nEn_grayColor} and \ref{fig:sEnKF_ETKF_based_GMM_minRMSE_vs_nEn_grayColor} (to be shown below), respectively, at different ensemble sizes.

\subsubsection{Effect of the Ensemble Size}

In the second experiment we examine the effect of the ensemble size in
each ensemble filter on the performance of the corresponding
PEnKF. For reference, we also examine the performance of the SEnKF and
the ETKF under various ensemble sizes. The experiment settings are the
same as those in \S~\ref{sec:lin_exp_n_vs_c}. In the PSEnKF and PETKF,
we choose the fraction coefficient $c \in \{0.05:0.1:0.95\}$. We also
choose the number of ensemble filters in each PEnKF to be $3$. For
each individual EnKF in the corresponding PEnKF, we let the ensemble
size $m$ take values from the set
$\{20,40,80,100,200,400,800,1000\}$, and for the experiments on the
single EnKF, we let $m \in
\{20,40,60,80,100,200,400,600,800,1000\}$. To conduct covariance
inflation and localization in each individual EnKF, we choose the
inflation factor $\delta=0.02$, and the length scale $l_c=50$. As in
\S~\ref{sec:lin_exp_n_vs_c}, covariance localization is conducted only
if the ensemble size $m$ is no larger than the dimension $40$.

\begin{figure*} %-- 8 --
\vspace*{2mm}
\centering

\includegraphics[width=0.8\textwidth]{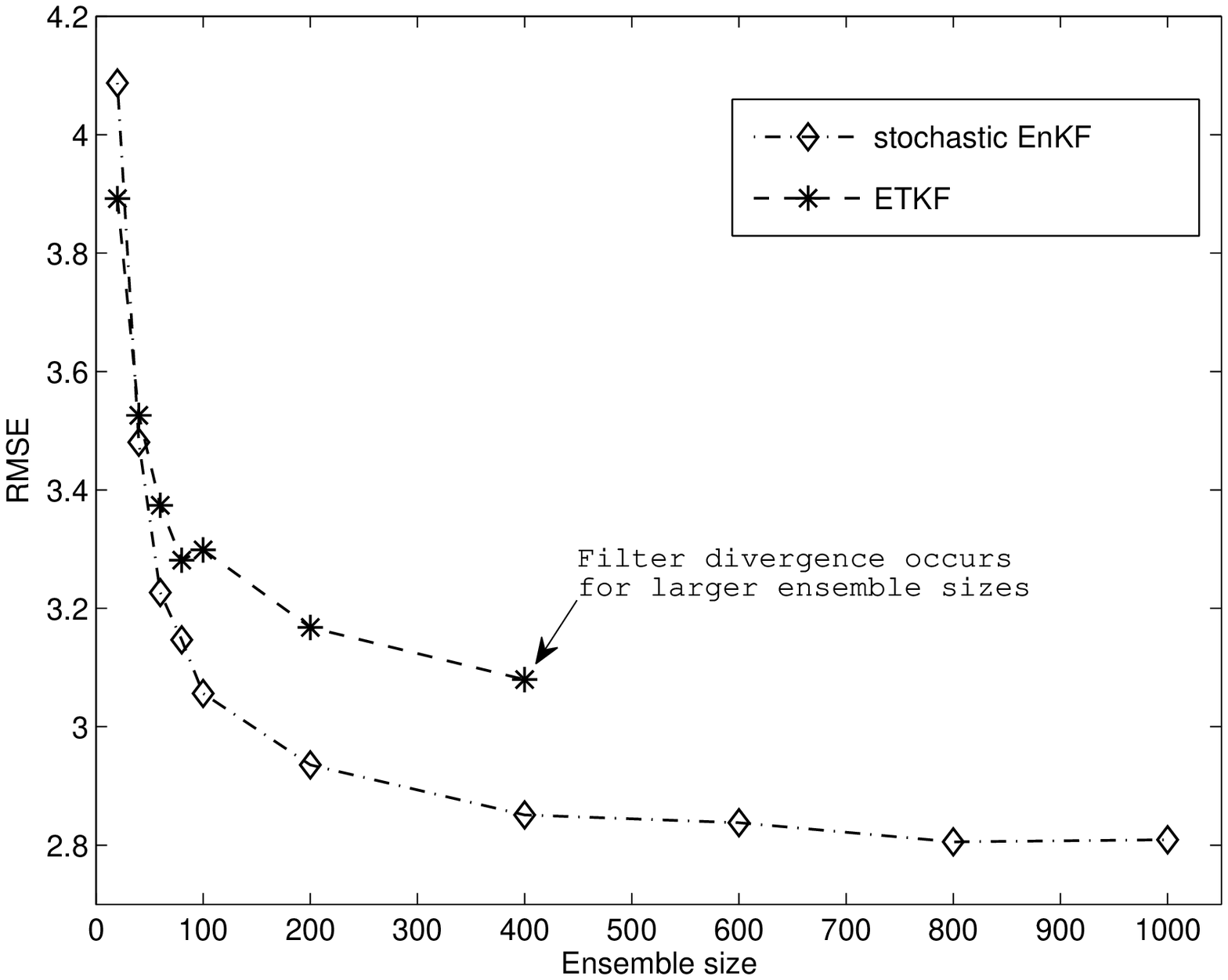}

\caption{\label{fig:sEnKF_ETKF_RMSE_vs_nEn_grayColor} RMS errors (over
  $20$ experiments) of the stochastic EnKF and the ETKF as the
  functions of the ensemble size.}
\end{figure*}

Fig.~\ref{fig:sEnKF_ETKF_RMSE_vs_nEn_grayColor} shows the rms errors
of the SEnKF and the ETKF as functions of the ensemble size $m$. For
both filters, their rms errors decrease as the ensemble size $m$
increases. The ETKF performs better than the SEnKF in the small sample
scenario with $m=20$. But as $m$ increases, the SEnKF outperforms the
ETKF instead. In particular, divergence in the ETKF occurs if $m>400$,
which did not happen in the linear observer case
(cf. Fig.~\ref{fig_lin:sEnKF_ETKF_RMSE_vs_nEn_grayColor}). On the
other hand, the rmse of the SEnKF appears to reach a plateau for
$m>400$, similar to the linear observer case. Comparing
Fig.~\ref{fig:sEnKF_ETKF_RMSE_vs_nEn_grayColor} with
Fig.~\ref{fig_lin:sEnKF_ETKF_RMSE_vs_nEn_grayColor}, it is easy to see
that, except for the stochastic EnKF at $m=20$, the presence of nonlinearity in the observer
deteriorates the performance of the ensemble filters.

\begin{figure*} % -- 9 --
\vspace*{2mm}
\centering
%--
	\subfigure[Stochastic EnKF-based PKF]{
	\label{sEnKF_based_GMM_contour_fraction_vs_nEn_grayColor}	
	\includegraphics[width=0.8\textwidth]
{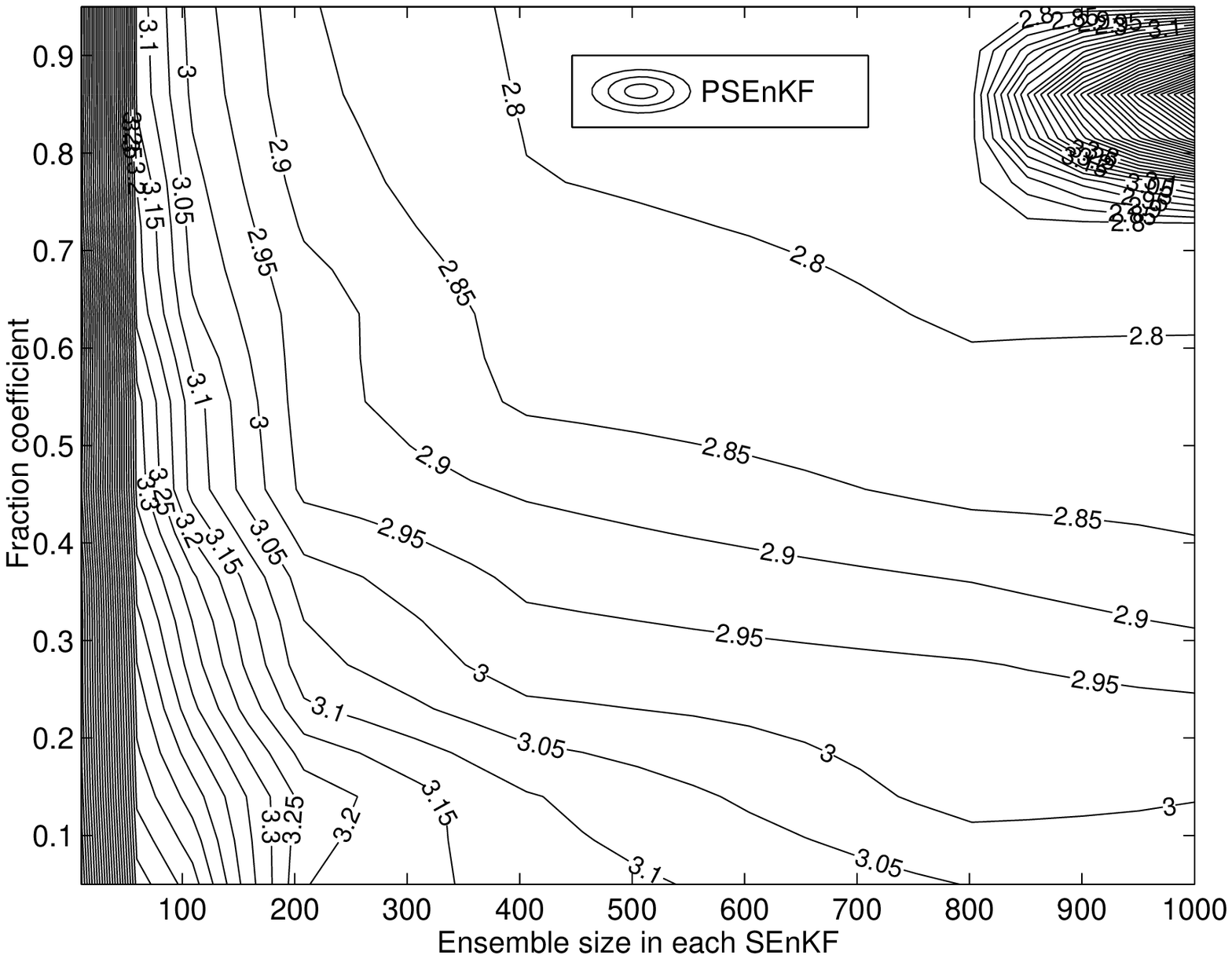}
     }
%--
	\subfigure[ETKF-based PKF]{
	\label{fig:ETKF_based_GMM_contour_fraction_vs_nEn_grayColor}	
	\includegraphics[width=0.8\textwidth]
{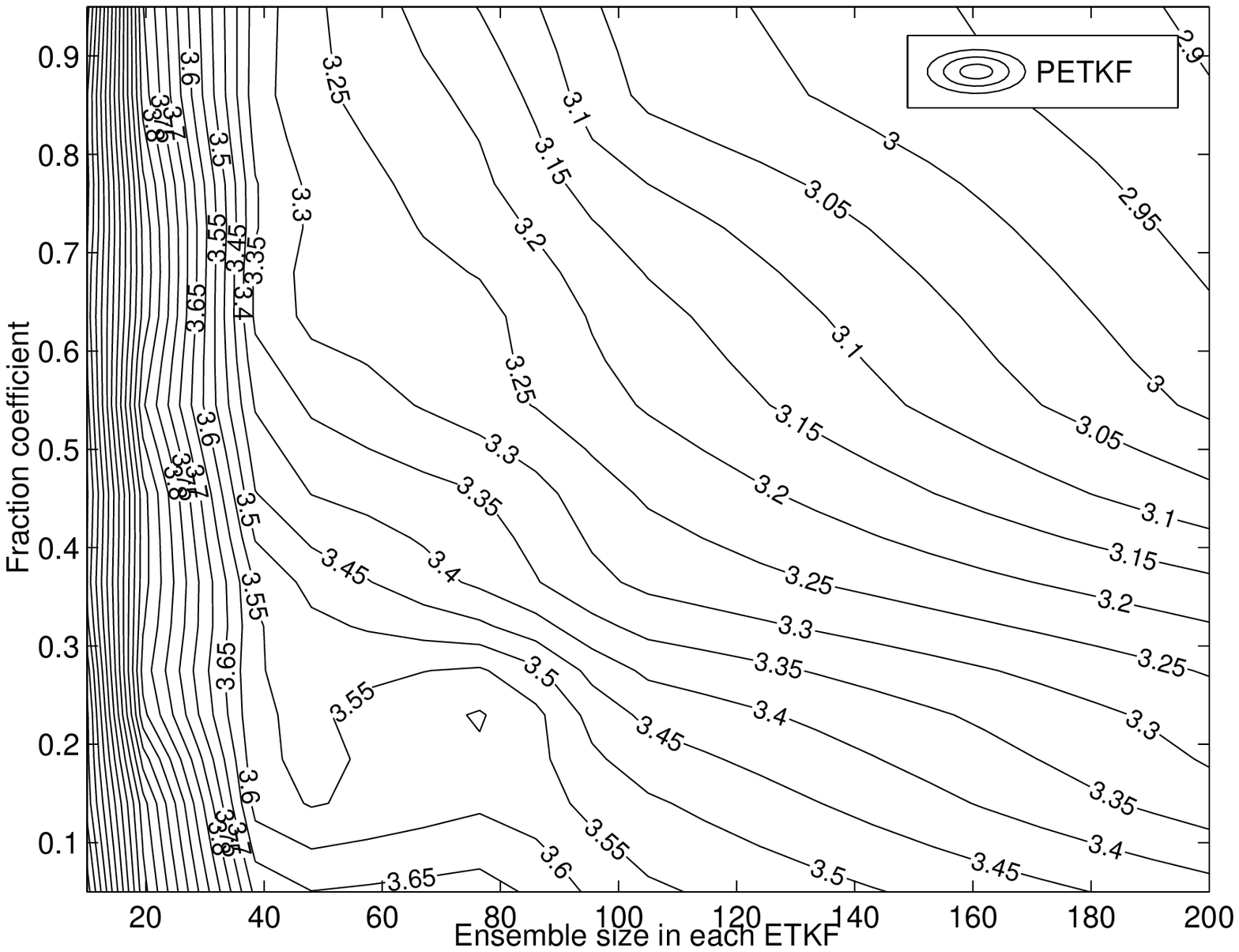}
     }
     \caption{\label{fig:GMM_contour_fraction_vs_nEn_grayColor} RMS
       errors (over $20$ experiments) of the stochastic EnKF- and
       ETKF-based PEnKFs (with a fixed number of Gaussian {\it pdf}s of $3$ in
       each PKF) as the functions of the fraction coefficient and the
       ensemble size of the ensemble filter. In
       Fig.~\ref{fig:ETKF_based_GMM_contour_fraction_vs_nEn_grayColor}
       the ensemble size in each ensemble filter is only up to
       $200$. Divergence occurs in the ETKF-based PKF with ensemble
       sizes in each ensemble filter larger than $200$.}
\end{figure*}

Fig.~\ref{fig:GMM_contour_fraction_vs_nEn_grayColor} plots the rms
errors of the PSEnKF and the PETKF as functions of the fraction
coefficient $c$, and the ensemble size $m$ in the corresponding SEnKF
and the ETKF, respectively. In the PSEnKF
(cf. Fig.~\ref{sEnKF_based_GMM_contour_fraction_vs_nEn_grayColor}),
the rmse tends to decrease as both $c$ and $m$ increases when the
ensemble size $m \leq 800$. However, when $m>800$, the impact of $m$
on the filter performance is not significant, which is consistent with
the results in Fig.~\ref{fig:sEnKF_ETKF_RMSE_vs_nEn_grayColor}. On the
other hand, in the PETKF
(cf. Fig.~\ref{fig:ETKF_based_GMM_contour_fraction_vs_nEn_grayColor}),
filter divergence occurs for $m>200$, which is why we only report its
rmse with $m \leq 200$ in
Fig.~\ref{fig:ETKF_based_GMM_contour_fraction_vs_nEn_grayColor}, where
the rmse of the PETKF appears to be a monotonically decreasing
function of $m$ and $c$.

\begin{figure*} %-- 10 --
\vspace*{2mm}
\centering

\includegraphics[width=0.8\textwidth]{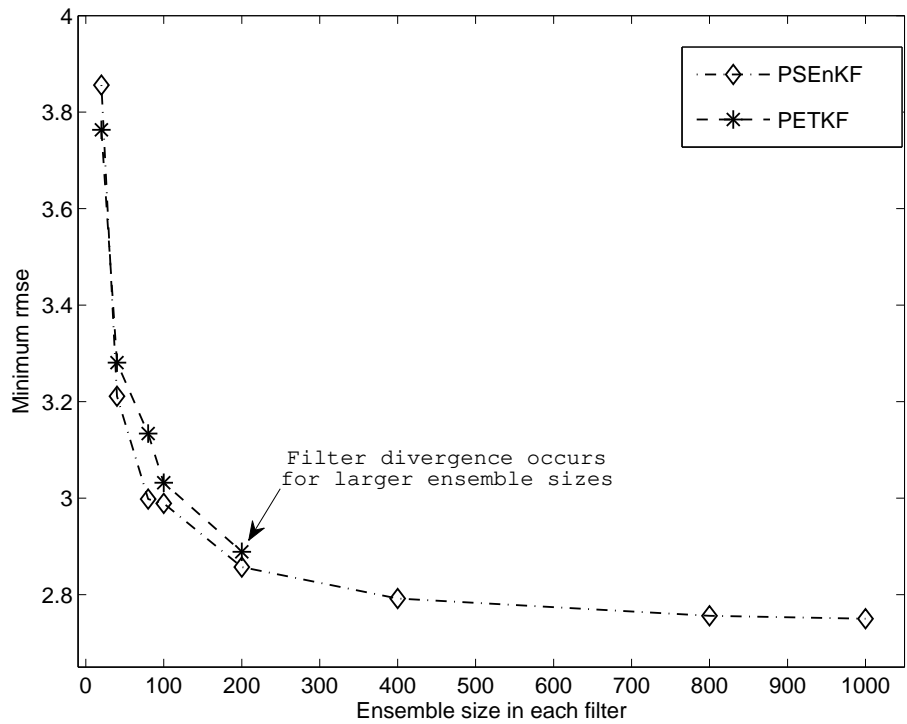}

\caption{\label{fig:sEnKF_ETKF_based_GMM_minRMSE_vs_nEn_grayColor}
  Minimum rms errors $\hat{e}_{min}$ (over $20$ experiments) of the
  stochastic EnKF- and ETKF-based PEnKFs (with a fixed number of
  Gaussian {\it pdf}s of $3$ in each PKF) as the function of the ensemble
  size in each ensemble filter.}
\end{figure*}

In analogy to the first experiment,
Fig.~\ref{fig:sEnKF_ETKF_based_GMM_minRMSE_vs_nEn_grayColor} plots the
minimum rms errors $\hat{e}_{min}$ of the PSEnKF and the PETKF within
the tested fraction coefficient $c$ and ensemble size $m$. One may
observe that, similar to the SEnKF and the ETKF themselves, the
$\hat{e}_{min}$ of both the PSEnKF and the PETKF decrease as $m$
increases. However, for the PETKF, divergence occurs if $m>200$,
rather than $m>400$ as in
Fig.~\ref{fig:sEnKF_ETKF_RMSE_vs_nEn_grayColor}, but overall its rmse
is closer to that obtained in the PSEnKF. Meanwhile, a comparison
between Fig.~\ref{fig:sEnKF_ETKF_RMSE_vs_nEn_grayColor} and
Fig.~\ref{fig:sEnKF_ETKF_based_GMM_minRMSE_vs_nEn_grayColor} shows that the PEnKFs perform better than the corresponding EnKFs. Also, a comparison between Fig. \ref{fig_lin:sEnKF_ETKF_based_GMM_minRMSE_vs_nEn_grayColor} and \ref{fig:sEnKF_ETKF_based_GMM_minRMSE_vs_nEn_grayColor} shows that, except for the PSEnKF at $m=20$, the nonlinearity in the observer again deteriorates the performance of the ensemble filters.

%##############################
%  DISCUSSION
%##############################

\section{Discussion}
\label{discussion}

This paper presented a discrete solution of the optimal nonlinear
filter, called the particle Kalman filter (PKF), based on the Gaussian
mixture representation of the state {\it pdf} given the
observations. The PKF solves the nonlinear Bayesian correction step by
complementing the Kalman filter-like correction step of the particles
with a particle filter-like correction step of the weights. The PKF
simultaneously runs a weighted ensemble of the Kalman filters in
parallel. This is far beyond our computing capabilities when dealing
with computationally demanding systems, as the atmospheric and oceanic
models. Therefore, to reduce computational cost, one may instead
consider a low-rank parametrization of the Gaussian mixture covariance
matrices of the state {\it pdf}s. An efficient way to do that is to
resort to the ensemble Kalman filter (EnKF) and use an EnKF-like
method to update each component of the Gaussian mixture {\it
  pdf}s. This amounts to running a weighted ensemble of the EnKFs. In
this work, the PKF was implemented using the stochastic EnKF and a
deterministic EnKF, the ensemble transform Kalman filter (ETKF). We
call this type of implementation the particle ensemble Kalman filter
(PEnKF).

The PEnKF sets a nonlinear Bayesian filtering framework that
encompasses the EnKF methods as a special case. As in the EnKF, the
Kalman correction in the PEnKF attenuates the degeneracy of the
ensemble by allocating the ensemble members far away from the incoming
observation relatively more weights than in the particle filter, so
that the filter can operate with reasonable size ensembles. To further
improve the performance of the PEnKF, we also introduced to the PEnKF
a re-sampling step similar to that used in the regularized particle
filter \citep{Musso-improving-2001,Stavropoulos-improved-2001}.

The stochastic EnKF and ETKF-based PEnKFs, called the PSEnKF and the
PETKF, respectively, were implemented and their performance was
investigated with the strongly nonlinear Lorenz-96 model. These
filters were tested with both linear and nonlinear observation
operators. Experiments results suggest that the PSEnKF and the PETKF
outperform their corresponding EnKFs. It was also found that the ETKF
outperforms the stochastic EnKF for small size ensembles while the
stochastic EnKF exhibits better performance for large size
ensembles. We argued that this happens because the EnKF endures less
observational sampling errors when the ensemble size is large. Another
reason would also be the better approximation of the PEnKF
distributions provided by the stochastic EnKF compared to the
ETKF. This was also true for their PEnKF counterparts. Overall, the
conclusions from the numerical results obtained with the linear and
nonlinear observation operators were not fundamentally different,
except that in general better estimation accuracy was achieved with the linear
observer when the sampling error is not the dominant factor.
The results also suggest that the PEnKFs could more benefit from the use of more components in the
mixture of normals (MON) and larger ensembles in the EnKFs in the
nonlinear observations case.

Future work will focus on developing and testing new variants of the
PEnKF that applies more efficient approximations, in term of
computational cost, to update the mixture covariance matrices. Another
direction for improvement would be also to work on localizing the
correction step of the particle weights \citep{VanLeeuwen2009}. Our
final goal is to develop a set of computationally feasible suboptimal
PEnKFs that can outperform the EnKF methods at reasonable
computational cost. As stated by \citet{Anderson2003-local},
developing filters in the context of the optimal nonlinear filtering
problem, rather than starting from the Kalman filter, should lead to a
more straightforward understanding of their capabilities.

The paper further discussed how the PEnKF can also be used as a
general framework to simultaneously run several assimilation
systems. We believe that this approach provides a framework to merge
the solutions of different EnKFs, or to develop hybrid
EnKF-variational methods. Work in this direction is under
investigation.

\section*{Acknowledge}
We would like to thank the three anonymous reviewers for their
valuable comments and suggestions. Ibrahim Hoteit was partially
supported by ONR grant N00014-08-1-0554.

%###########
%  APPENDIX
%###########

%\appendix
%\appendixpage

%##############################
%  BIBLIOGRAPHY
%##############################

\bibliographystyle{ametsoc}
\bibliography{references}

%##############################
%  FIGURES
%#############################

\newpage

\listoffigures

\clearpage
%\section*{\centering Figure Captions}

\newpage

% Figure-1: Domain.
%~\vspace{2cm}\\
%\begin{figure}[ht]
%  \centering
%  \includegraphics[height = 7cm, width = 12cm]{Figures/fig1}
%  \caption{Topography of the $1/4^\circ$ Mediterranean model.}
%  \label{fig:domain}
%\end{figure}

%\newpage

% Figure-2: Main Experiment - RRMS.
%~\vspace{2cm}\\
%\begin{figure}[ht]
%  \centering
%  \includegraphics[height = 7cm, width = 12cm]{Figures/fig2}
%  \caption{Time evolution of the basin average kinetic energy ($10{^-3}m^2/s^2$) from $1979$ to $1987$.}
%  \label{fig:energy}
%\end{figure}

%\newpage

%-------------------------------------------LINEAR CASE----------------------------------------------------------
\clearpage

\clearpage

\clearpage

\clearpage

\clearpage

%-------------------------------------------NONLINEAR CASE----------------------------------------------------------
\clearpage

\clearpage

\clearpage

\clearpage

\clearpage

%%%%%%%%%%%%%%%%%%%%%%%%%%%%%%%%%%%%%%%%
% SUPPORT MATERIALS %
\clearpage
\setcounter{page}{1}
\setcounter{equation}{0}
\renewcommand{\theequation}{S.\arabic{equation}}
\section*{\centering Support Material: The Full Re-sampling Algorithm}
\label{resamp}

Here we discuss how to construct the ensemble set $\{
\mathbf{X}_{en}^i, i=1,\dotsb,q\}$ in the PEnKF. We note that the
relative positions of the dimension $n$ of the random vector
$\mathbf{x}$, the number $q$ of the Gaussian {\it pdf}s in the MON
Eq.~(\ref{eq:new_GMM}), and the ensemble size $m$ of each EnKF in the
PKF determines our re-sampling strategies. In certain circumstances, a
singular value decomposition (SVD) may be required on the covariance
matrix $\bar{\mathbf{P}}$ in Eq.~(\ref{eq:stat_original_pdf}) such
that
\begin{equation}
  \label{eq:SVD_on_P}
  \bar{\mathbf{P}} = \mathbf{V}
  \mathbf{D} \mathbf{V}^T = \sum\limits_{i=1}^{n} \sigma_i^2
  \mathbf{e}_i \mathbf{e}_i^T \, ,
\end{equation}
where $\mathbf{V}$ is the matrix consisting of the eigenvectors
$\mathbf{e}_i$ of $\bar{P}$, and $\mathbf{D} \equiv \text{diag}
(\sigma_1^2,\dotsb,\sigma_n^2)$ the diagonal matrix consisting of the
corresponding eigenvalues $\sigma_i^2$ (we also assume $\sigma_i \geq
0$ without loss of generality). Depending on the values of $q$, $m$
and $n$, one may avoid computing the full spectra of
$\bar{\mathbf{P}}$, as to be shown below.

%%%%%%%%%%%%%%%%%%%%%%%%%%%%%%%%%%%%%%%
\newcounter{caseCounter}
\setcounter{caseCounter}{1}
%--
\subsection*{Case \Roman{caseCounter}: $q \leq n$ and $m \leq n$}
%%%%%%%%%%%%%%%%%%%%%%%%%%%%%%%%%%%%%%%

In this case the number $q$ of (re-approximation) Gaussian
distributions and the ensemble size $m$ are both less than the
dimension $n$ of the system state. We consider two possibilities
below.

\subsubsection*{$1.~q \leq m \leq n$}
Here we choose
\begin{subequations}
\begin{align}
\label{eq:case1_center_cov} & \dfrac{1}{q} \sum_{i=1}^{q} \left(  \mathbf{\theta}_i - \bar{\mathbf{x}} \right) \left(  \mathbf{\theta}_i - \bar{\mathbf{x}}  \right)^T = (1-c^2) \sum\limits_{i=1}^{q-1} \sigma_i^2 \mathbf{e}_i \mathbf{e}_i^T \, , \\
\label{eq:case1_ensemble_cov}  & \mathbf{\Phi}  = c^2 \sum\limits_{i=1}^{q-1} \sigma_i^2 \mathbf{e}_i \mathbf{e}_i^T + \sum\limits_{i=q}^{m-1} \sigma_i^2 \mathbf{e}_i \mathbf{e}_i^T \, .
\end{align}
\end{subequations}
%Following \citet{Luo2008-spgsf1}, we discuss the influence of $c$ as follows. When $c \rightarrow 0$, we have $\mathbf{\Phi} \rightarrow \mathbf{0}$, given that the eigenvalues $\sigma_i^2 \rightarrow 0$ for sufficiently large $N$. In this case, $\tilde{p} \left( \mathbf{x} \right)$ in Eq.~(\ref{eq:new_GMM}) approaches the Monte Carlo approximation as used in the particle filter, with the mass points equal to $\mathbf{\theta}_i$. On the other hand, when $c \rightarrow 1$, we have $\mathbf{\Phi}$ approach the first $(n-1)$ leading modes $\sigma_i^2 \mathbf{e}_i \mathbf{e}_i^T$ of $\bar{\mathbf{P}}$ (in particular, $\mathbf{\Phi} \rightarrow \bar{\mathbf{P}}$ itself when $n$ is close to $n$), while $\mathbf{\theta}_i$ in Eq.~(\ref{eq:case1_center_cov}) approaches $\bar{\mathbf{x}}$ for all $i$. As a result, $\tilde{p} \left( \mathbf{x} \right)$ in Eq.~(\ref{eq:new_GMM}) (approximately) approaches the Gaussian pdf $N(\mathbf{x}: \bar{\mathbf{x}},\bar{\mathbf{P}})$, which is essentially the assumption used in the EnKF. In this sense, when equipped with the re-sampling procedures, the PEnKF is a filter in between the particle filter and the EnKF, with an adjustable parameter $c$ that influences its behavior. Similar observations on the influence of $c$ can be obtained in the other cases.
The reason to choose the superscripts $q-1$ and $m-1$ on the right
hand side of Eqs.~(\ref{eq:case1_center_cov}) and
(\ref{eq:case1_ensemble_cov}) will be made clear soon. We also note
that the sum
\begin{equation}
  \mathbf{\Phi} + \dfrac{1}{q} \sum_{i=1}^{q} \left(  \mathbf{\theta}_i - \bar{\mathbf{x}} \right) \left(  \mathbf{\theta}_i - \bar{\mathbf{x}}  \right)^T = \sum\limits_{i=1}^{n-1} \sigma_i^2 \mathbf{e}_i \mathbf{e}_i^T
\end{equation}
is not equal to $\bar{\mathbf{P}}$ exactly. Instead, it only adds up
to the first $(m-1)$ terms of $\sigma_i^2 \mathbf{e}_i
\mathbf{e}_i^T$.

Let $\mathbf{\Theta} = [\mathbf{\theta}_1, \dotsb, \mathbf{\theta}_q]$
be the collection of the means $\mathbf{\theta}_i$ in the MON
$\tilde{p} \left( \mathbf{x} \right)$, and
\[
\mathbf{S}_{\mu}= \sqrt{1-c^2} \, [\sigma_1 \mathbf{e}_1, \dotsb, \sigma_{q-1} \mathbf{e}_{q-1}]
\]
be the square root of $(1-c^2) \sum\limits_{i=1}^{q-1} \sigma_i^2
\mathbf{e}_i \mathbf{e}_i^T$ in Eq.~(\ref{eq:case1_center_cov}), then
it can be verified that
\begin{equation} \label{eq:case1_mean_generation}
\mathbf{\Theta} = \bar{\mathbf{x}} \, \mathbf{1}_q^T + \sqrt{q} \, \mathbf{S}_{\mu}\mathbf{C}_{q-1,q}
\end{equation}
yields a set of the means $\mathbf{\theta}_i$ that satisfy
Eq.~(\ref{eq:case1_center_cov}), where $\mathbf{1}_q^T$ denotes the
transpose of the $q \times 1$ column vector $\mathbf{1}_q$ with all
its elements being one (so that $\bar{\mathbf{x}} \, \mathbf{1}_q^T$
consists of $N$ identical column vectors $\bar{\mathbf{x}}$), and
$\mathbf{C}_{q-1,q}$ is a $(q-1) \times q$ matrix satisfying that
$\mathbf{C}_{q-1,q} (\mathbf{C}_{q-1,q})^T = \mathbf{I}_{q-1}$, with
$\mathbf{I}_{q-1}$ being the $(q-1)$-dimensional identity matrix, and
that $\mathbf{C}_{q-1,q} \mathbf{1}_q = \mathbf{0}_{q-1}$, with
$\mathbf{0}_{q-1}$ being a $(q-1) \times 1$ column vector with all its
elements being zero. The first constraint, $\mathbf{C}_{q-1,q}
(\mathbf{C}_{q-1,q})^T = \mathbf{I}_{q-1}$ guarantees that the sample
covariance of $\mathbf{\Theta}$ satisfies the constraint in
Eq.~(\ref{eq:case1_center_cov}), and the second one,
$\mathbf{C}_{q-1,q} \mathbf{1}_q = \mathbf{0}_{q-1}$ guarantees that
the sample mean of $\mathbf{\Theta}$ is equal to $\bar{\mathbf{x}}$,
as is required in Eq.~(\ref{eq:objective_expression}). For the
generation of such a matrix $\mathbf{C}_{q-1,q}$, readers are referred
to, for example, \citet{Hoteit2002,Pham2001}. In addition, since the
dimension of $\mathbf{C}_{q-1,q}$ is $(q-1) \times q$, we require that
the dimension of the square root matrix $\mathbf{S}_{\mu}$ is $n
\times (q-1)$. Therefore, on the right hand side of
Eq.~(\ref{eq:case1_center_cov}), the superscript shall be $(q-1)$,
rather than $q$. The reason to use the superscript $(m-1)$ in
Eq.~(\ref{eq:case1_ensemble_cov}) is similar, as can be seen below.

To generate the ensembles $\mathbf{X}_{en}^i$ ($i=1,\dotsb,q$), with $\mathbf{\theta}_i$ and $\mathbf{\Phi}$ being their sample means and covariances, we first construct the square root matrix
\begin{equation}
\mathbf{S}_{\phi} = [c \sigma_1 \mathbf{e}_1, \dotsb, c \sigma_{q-1} \mathbf{e}_{q-1}, \sigma_{q} \mathbf{e}_{q}, \dotsb, \sigma_{n-1} \mathbf{e}_{m-1}]
\end{equation}
of $\mathbf{\Phi}$, and generate $\mathbf{X}_{en}^i$ by
\begin{equation} \label{eq:case1_ensemble_generation}
\mathbf{X}_{en}^i = \mathbf{\theta}_i \, \mathbf{1}_m^T + \sqrt{m} \, \mathbf{S}_{\phi} \mathbf{C}_{m-1,m} \, ,~\text{for}~i=1,\dotsb, q \, ,
\end{equation}
where $\mathbf{C}_{m-1,m}$ is a matrix similar to $\mathbf{C}_{q-1,q}$
in Eq.~(\ref{eq:case1_mean_generation}). We note that the term
$\sqrt{m} \, \mathbf{S}_{\phi} \mathbf{C}_{m-1,m}$ is common to all
EnKFs, and thus only needs to be calculated once. This is direct
implication from the choice of the uniform covariance $\mathbf{\Phi}$
in $ \tilde{p} \left( \mathbf{x} \right)$, as we have pointed out
previously, which leads to computational savings in comparison to the
non-uniform choice.

\subsubsection*{$2.~m < q \leq n$}
Here we choose
\begin{subequations}
\begin{align}
\label{eq:case12_center_cov} & \dfrac{1}{q} \sum_{i=1}^{q} \left(  \mathbf{\theta}_i - \bar{\mathbf{x}} \right) \left(  \mathbf{\theta}_i - \bar{\mathbf{x}}  \right)^T = (1-c^2) \sum\limits_{i=1}^{m-1} \sigma_i^2 \mathbf{e}_i \mathbf{e}_i^T + \sum\limits_{i=m}^{q-1} \sigma_i^2 \mathbf{e}_i \mathbf{e}_i^T \, , \\
\label{eq:case12_ensemble_cov}  & \mathbf{\Phi}  = c^2 \sum\limits_{i=1}^{m-1} \sigma_i^2 \mathbf{e}_i \mathbf{e}_i^T \, .
\end{align}
\end{subequations}

Now define the square root matrix
\begin{equation}
  \mathbf{S}_{\mu} = [\sqrt{1-c^2} \sigma_1 \mathbf{e}_1,\dotsb,\sqrt{1-c^2} \sigma_{m-1} \mathbf{e}_{m-1},\sigma_{m} \mathbf{e}_{m},\dotsb,\sigma_{q-1} \mathbf{e}_{q-1}]
\end{equation}
of the term on right hand side of Eq.~(\ref{eq:case12_center_cov}),
and the square root matrix
\begin{equation}
\mathbf{S}_{\phi} = c \, [ \sigma_1 \mathbf{e}_1,\dotsb, \sigma_{n-1} \mathbf{e}_{m-1}]
\end{equation}
of $\mathbf{\Phi}$ in Eq.~(\ref{eq:case12_ensemble_cov}). Then
$\mathbf{\theta}_i$ and $\mathbf{X}_{en}^i$ can be generated through
Eqs.~(\ref{eq:case1_mean_generation}) and
(\ref{eq:case1_ensemble_generation}), respectively.

%%%%%%%%%%%%%%%%%%%%%%%%%%%%%%%%%%%%%%%
\addtocounter{caseCounter}{1}
%--
\subsection*{Case \Roman{caseCounter}: $q \leq n$ and $m > n$}
%%%%%%%%%%%%%%%%%%%%%%%%%%%%%%%%%%%%%%%
In this case the number $q$ of Gaussian distributions is less than the
dimension $n$ of the system state, but the ensemble size $m$ is larger
than $n$. We choose
\begin{subequations}
\begin{align}
\label{eq:case2_center_cov} & \dfrac{1}{q} \sum_{i=1}^{q} \left(  \mathbf{\theta}_i - \bar{\mathbf{x}} \right) \left(  \mathbf{\theta}_i - \bar{\mathbf{x}}  \right)^T = (1-c^2) \sum\limits_{i=1}^{q-1} \sigma_i^2 \mathbf{e}_i \mathbf{e}_i^T \, , \\
\label{eq:case2_ensemble_cov}  & \mathbf{\Phi}  = c^2 \sum\limits_{i=1}^{q-1} \sigma_i^2 \mathbf{e}_i \mathbf{e}_i^T + \sum\limits_{i=q}^{n} \sigma_i^2 \mathbf{e}_i \mathbf{e}_i^T = \bar{\mathbf{P}} - (1-c^2) \sum\limits_{i=1}^{q-1} \sigma_i^2 \mathbf{e}_i \mathbf{e}_i^T \, .
\end{align}
\end{subequations}
The last equality in Eq.~(\ref{eq:case2_ensemble_cov}) implies that one does not need to compute the full spectra of $\bar{\mathbf{P}}$ and the corresponding eigenvectors. Instead, one only needs to compute the first $(q-1)$ terms of $\sigma_i^2 \mathbf{e}_i \mathbf{e}_i^T$.

Now define the square root matrix
\begin{equation}
\mathbf{S}_{\mu} = \sqrt{1-c^2} [\sigma_1 \mathbf{e}_1,\dotsb, \sigma_{q-1} \mathbf{e}_{q-1}] \, ,
\end{equation}
so that one can again adopt Eq.~(\ref{eq:case1_mean_generation}) to
generate $\mathbf{\theta}_i$ ($i=1,\dotsb,q$). To generate the
ensembles $\mathbf{X}_{en}^i$, the situation here is different from
that in the previous case, in that the ensemble size $m$ is larger
than the dimension $n$, so that one cannot obtain enough ensemble
members through Eq.~(\ref{eq:case1_ensemble_generation}). As a result,
one may instead choose to draw $(m-1)$ samples $\delta
\mathbf{x}_{j}^{\phi}$ ($j=1,\dotsb,m-1$) from the distribution
$N(\delta \mathbf{x}: \mathbf{0}_n,\mathbf{\Phi})$ to form a matrix
$\Delta \mathbf{X}_{\phi} \equiv [ \delta
\mathbf{x}_{1}^{\phi},\dotsb, \delta \mathbf{x}_{m-1}^{\phi}]$. Then
the ensemble $\mathbf{X}_{en}^i$ is produced via
\begin{equation} \label{eq:case2_ensemble_generation}
\mathbf{X}_{en}^i = \mathbf{\theta}_i \, \mathbf{1}_m^T + \Delta \mathbf{X}_{\phi} \mathbf{C}_{m-1,m}, \, ~\text{for}~i=1,\dotsb,q \, .
\end{equation}
Eq.~(\ref{eq:case2_ensemble_generation}) is similar to the partial
re-sampling scheme in \citet{Hoteit2008}, although here the
perturbation term $\Delta \mathbf{X}_{\phi} \mathbf{C}_{m-1,m}$ can be
common to all EnKFs, and thus can be drawn only once to reduce
computational
cost. %In general, drawing samples from a pdf incurs certain sampling errors, therefore we always intend to avoid using it if possible.

%%%%%%%%%%%%%%%%%%%%%%%%%%%%%%%%%%%%%%%
\addtocounter{caseCounter}{1}
%--
\subsection*{Case \Roman{caseCounter}: $q > n$ and $m \leq n$}
%%%%%%%%%%%%%%%%%%%%%%%%%%%%%%%%%%%%%%%
In this case the ensemble size $m$ is no larger than the dimension $n$
of the system state, but the number $q$ of Gaussian distributions
is. We choose
\begin{subequations}
\begin{align}
\label{eq:case3_center_cov} & \dfrac{1}{q} \sum_{i=1}^{q} \left(  \mathbf{\theta}_i - \bar{\mathbf{x}} \right) \left(  \mathbf{\theta}_i - \bar{\mathbf{x}}  \right)^T = (1-c^2) \sum\limits_{i=1}^{m-1} \sigma_i^2 \mathbf{e}_i \mathbf{e}_i^T + \sum\limits_{i=m}^{n} \sigma_i^2 \mathbf{e}_i \mathbf{e}_i^T = \bar{\mathbf{P}} - c^2 \sum\limits_{i=1}^{m-1} \sigma_i^2 \mathbf{e}_i \mathbf{e}_i^T \, , \\
\label{eq:case3_ensemble_cov}  & \mathbf{\Phi}  = c^2 \sum\limits_{i=1}^{m-1} \sigma_i^2 \mathbf{e}_i \mathbf{e}_i^T  \, .
\end{align}
\end{subequations}
Since $q > n$, we choose to draw $(q-1)$ samples $\delta
\mathbf{x}_j^{\mu}$ from the distribution $N(\delta \mathbf{x}:
\mathbf{0}_n,\bar{\mathbf{P}} - c^2 \sum\limits_{i=1}^{m-1} \sigma_i^2
\mathbf{e}_i \mathbf{e}_i^T)$ to form a matrix $\Delta
\mathbf{X}_{\mu} \equiv [ \delta \mathbf{x}_{1}^{\mu}, \dotsb, \delta
\mathbf{x}_{q-1}^{\mu}]$, while $\mathbf{\theta}_i$ are generated by
\begin{equation} \label{eq:case3_mean_generation}
\mathbf{\Theta} = \bar{\mathbf{x}} \, \mathbf{1}_q^T + \Delta \mathbf{X}_{\mu} \mathbf{C}_{q-1,q} \, .
\end{equation}

Let
\begin{equation}
\mathbf{S}_{\phi} = c [\sigma_1 \mathbf{e}_1,\dotsb, \sigma_{n-1} \mathbf{e}_{m-1}] \, ,
\end{equation}
then $\mathbf{X}_{en}^i$ can be generated through
Eq.~(\ref{eq:case1_ensemble_generation}).

%%%%%%%%%%%%%%%%%%%%%%%%%%%%%%%%%%%%%%%
\addtocounter{caseCounter}{1}
%--
\subsection*{Case \Roman{caseCounter}: $q > n$ and $m > n$}
%%%%%%%%%%%%%%%%%%%%%%%%%%%%%%%%%%%%%%%
In this case both the number $q$ of Gaussian distributions and the
ensemble size $m$ are larger than the dimension $n$ of the system
state. We let $\mathbf{\Phi} = c^2 \bar{\mathbf{P}}$ and define
$\mathbf{P}_n = (1-c^2) \bar{\mathbf{P}}$. To generate
$\mathbf{\theta}_i$, we first draw $(q-1)$ samples $\delta
\mathbf{x}_j^{\mu}$ from the distribution $N(\delta \mathbf{x}:
\mathbf{0}_n,\mathbf{P}_n)$ to form a matrix $\Delta \mathbf{X}_{\mu}
= [ \delta \mathbf{x}_{1}^{\mu}, \dotsb, \delta
\mathbf{x}_{q-1}^{\mu}]$, and then apply
Eq.~(\ref{eq:case3_mean_generation}). Meanwhile, we also draw $(m-1)$
samples $\delta \mathbf{x}_{j}^{\phi}$ from the distribution $N(\delta
\mathbf{x}: \mathbf{0}_n,\mathbf{\Phi})$ to form a matrix $\Delta
\mathbf{X}_{\phi} \equiv [ \delta \mathbf{x}_{1}^{\phi},\dotsb,\delta
\mathbf{x}_{m-1}^{\phi}]$, and then apply
Eq.~(\ref{eq:case2_ensemble_generation}) to generate the ensembles
$\mathbf{X}_{en}^i$.

\end{document}